\documentclass[iop]{emulateapj}
 \usepackage{color}
 \usepackage{natbib}

\begin{document}

 \shorttitle{HD~141399}
 \title{A 4-Planet System Orbiting The K0V Star HD~141399}

 \author{Steven S. Vogt\altaffilmark{1}, R. Paul Butler\altaffilmark{2}, Eugenio J. Rivera\altaffilmark{1}, Robert Kibrick\altaffilmark{1}, Jennifer Burt\altaffilmark{1}, Russell Hanson\altaffilmark{1}, Stefano Meschiari\altaffilmark{3}, Gregory W. Henry\altaffilmark{4}, and Gregory Laughlin\altaffilmark{1}}

 \shortauthors{Vogt et. al}

 \altaffiltext{1}{UCO/Lick Observatory, Department of Astronomy and Astrophysics, University of California at Santa Cruz,Santa Cruz, CA 95064}
 \altaffiltext{2}{Department of Terrestrial Magnetism, Carnegie Institute of Washington, Washington, DC 20015}
 \altaffiltext{3}{McDonald Observatory, University of Texas at Austin, Austin, TX 78752}
 \altaffiltext{4}{Center of Excellence in Information Systems,
 Tennessee State University, Nashville, TN 37209}

 \begin{abstract}
 We present precision radial velocity (RV) datasets from Keck-HIRES and from Lick Observatory's new Automated Planet Finder Telescope and Levy Spectrometer on Mt. Hamilton that reveal a multiple-planet system orbiting the nearby, slightly evolved, K-type star HD~141399. Our 91 observations over 10.5 years suggest the presence of four planets with orbital periods of 94.35, 202.08, 1070.35,  and 3717.35 days and minimum masses of 0.46, 1.36, 1.22,  and 0.69 $M_J$ respectively. The orbital eccentricities of the three inner planets are small, and the phase curves are well sampled. The inner two planets lie just outside the 2:1 resonance, suggesting that the system may have experienced dissipative evolution during the protoplanetary disk phase. The fourth companion is a Jupiter-like planet with a Jupiter-like orbital period. Its orbital eccentricity is consistent with zero, but more data will be required for an accurate eccentricity determination.

 \end{abstract}


 \section{Introduction}

 The detection and characterization of extrasolar planets constitutes one of the high water marks for the entire field of astronomy and astrophysics. Slightly more than two decades ago, our own solar system was the only planetary system known, whereas at conservative last count, over a thousand extrasolar planets have now been confirmed, and thousands of additional high-quality planetary candidates are awaiting follow-up. The quest for Earth-mass and Earth-sized worlds has been effectively fulfilled, both by the Kepler Mission \citep{Borucki12} and by the Doppler velocity surveys \citep{Dumusque12}.

\begin{figure}
 \plotone{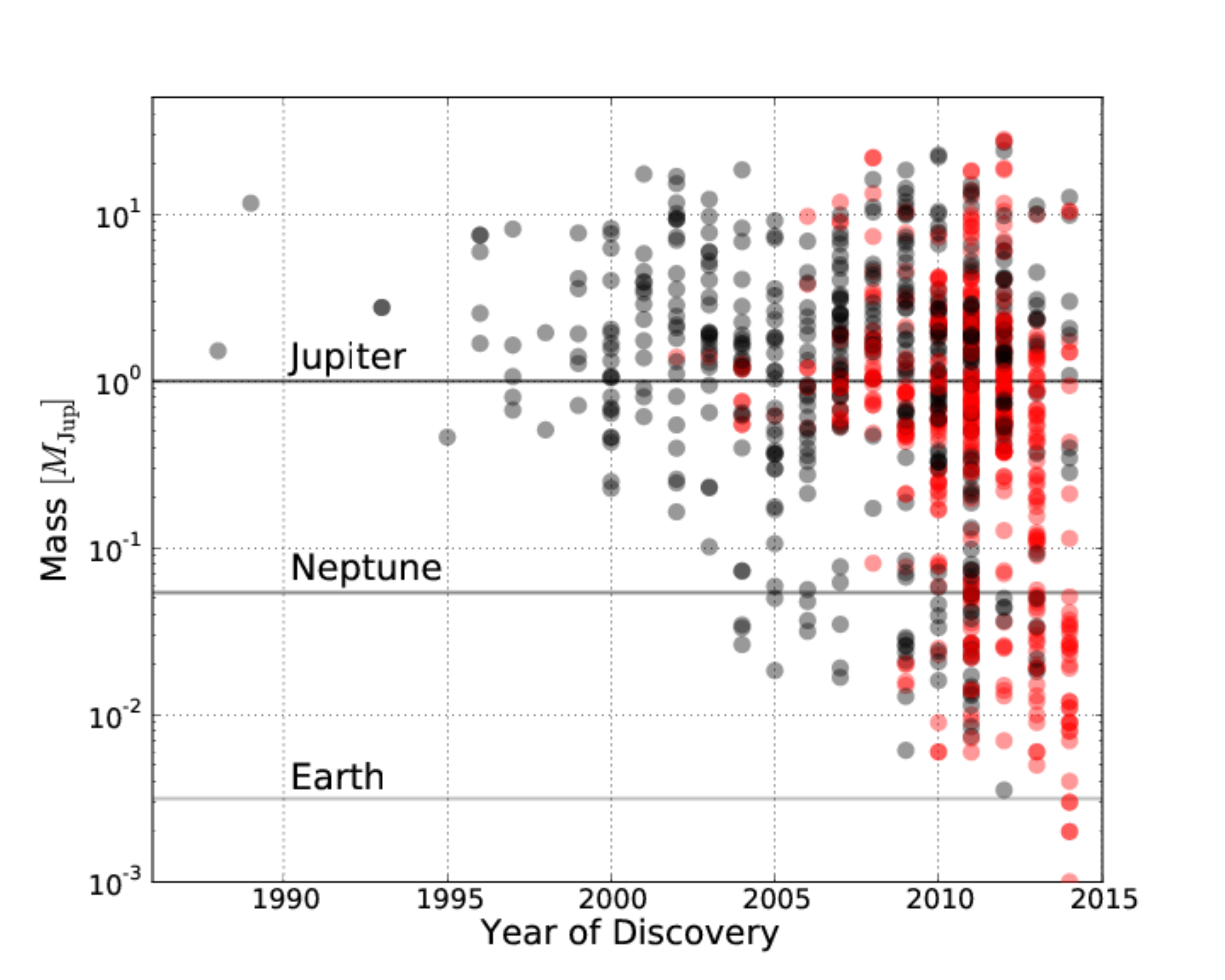}
 \caption{Diagram showing the masses of extrasolar planets versus their discovery year. Black points are extrasolar planets discovered by radial velocity observations with or without transits. Red points are extrasolar planets initially discovered by transits. The number of planets discovered by RV monitoring and having $M > 1.0\,M_J$ and $100 < P < 1000$ days has decreased in recent years, from an average of 18 such planets per year in 2007-2011, to 15 in 2012 and 7 in 2013. Data are from www.exoplanets.org and exoplanetarchive.ipac.caltech.edu, accessed 03/03/2013.}
 \label{fig:massYrDisc}
 \end{figure}

 The extraordinary string of successes has increasingly left the study of extrasolar planets standing at a crossroads. It has been clear for a number of years that the discovery of new planets merely for the sake of discovery has essentially lost its luster. Doppler velocity resources are increasingly being employed in the service of detecting very low mass planets orbiting very nearby stars \citep{Vogt10, Vogt12, Pepe11}, as well as in efforts to follow up on interesting transiting planetary candidates, especially those discovered by Kepler \citep{Howard13, Pepe13}. As a consequence, the steady stream of ``ordinary'' Jovian-mass planets with periods ranging from 100 to 1000 days orbiting nearby stars, as evidenced by Figure \ref{fig:massYrDisc}, has slowed to a relative trickle.

 The present article continues an extensive series of papers that have described the detection of the planetary systems that have emerged from the long-running Doppler Velocity monitoring efforts carried out with the Keck Telescope. Precision Doppler velocities were first obtained in quantity with Keck I's HIRES Spectrometer \citep{Vogt94} in 1996, following the installation of an iodine cell that can be inserted into the beam of collected starlight \citep{Butler96}. In the intervening years, a database of over 34,000 precision velocity measurements have been collected at Keck from a total of 1300 (mostly nearby) FGKM stars. The median time baseline over which velocities have been obtained for a given star in this catalog is $\tau_{d}$ = 2,563 days, and 197 stars have at least 50 high-precision radial velocity observations. Well over 100 planets have been detected with Keck, including ``firsts'', such as Gliese 436~b \citep{Butler04}, the first Neptune-mass extrasolar planet, Gliese 876~d, the first super-Earth with a secure mass determination \citep{Rivera05}, and HD~209458 \citep{Henry00}, the first known transiting planet.

 The target star considered in this paper, HD 141399, is a relatively nearby (36 pc distant) slightly evolved, slightly metal-rich K-type star located high in the northern sky ($46^\circ$ declination). Although it is quite bright, with V=7.2, its overall mediocrity has ensured that it has remained generally obscure, even in astronomical circles. A standard Simbad search, for example, turns up only four noncommittal mentions of the star in the literature between 1850 and 2013. Yet because HD 141399 is bright, and because it is chromospherically inactive, it has been on the Keck Radial Velocity program for over a decade. Its first iodine spectrum dates to July 2003. In recent months, it has also been repeatedly observed with the new Automated Planet Finder Telescope (APF) at Lick Observatory \citep{Vogt13}.

 In this paper, we report that our set of 91 velocities for HD~141399 (including 77 measurements obtained at Keck and 14 new measurements obtained at the APF) indicate that the star is accompanied by an unusual subsystem of three giant planets with near-circular orbits and with periods ranging from 94.0 days to 1070.0 days. The size of the annulus around this star spanned by these planetary orbits is associated with the zone of the terrestrial planets in our own solar system. Our radial velocities, furthermore, indicate the presence of a nearly Jupiter-mass planet at a Jupiter-like distance from the star with an orbital period of roughly a decade. Our time baseline of observations does not yet allow a definitive eccentricity determiniation for this outer planet. HD~141399's planetary system rises above a minimal threshold of interest as a consequence both of the proximity of its inner two planets to the 2:1 mean motion resonance, as well as the fact that it may well harbor an (apparently) rare near-twin of Jupiter.

 The plan for this paper is as follows: in \S 2 we review our Doppler velocity pipeline, with an emphasis on the new APF telescope. In \S 3, we give a brief overview of the properties of HD~141399. In \S 4, we discuss our four-planet model to explain the observed Doppler velocity variations exhibited by the star. In \S 5 we discuss the dynamical properties of the multiple-planet system that we have detected. In \S 6 we present our photometric observations of HD~141399 and in \S 7 we place our results into a larger context and conclude.

 \section{Radial Velocity Observations of HD~141399}

 Two platforms, the HIRES spectrometer \citep{Vogt94} of the Keck-I telescope, and the Levy spectrometer of the new Automated Planet Finder at Lick Observatory \citep{Vogt13} were used to obtain radial velocity measurements for HD~141399. Doppler shifts were measured in each case by placing an iodine absorption cell just ahead of the spectrometer slit in the converging beam of stellar light from the telescope \citep{1996PASP..108..500B}. The forest of iodine lines superimposed on the stellar spectra generates a wavelength calibration and enables measurement of each spectrometer's point spread function. The radial velocities from Keck were obtained by operating HIRES at a spectral resolving power $R\sim$70,000 over the wavelength range of 3700-8000 $\AA$, though only the region 5000-6200 $\AA$ containing a significant density of iodine lines was used in the present Doppler analysis. The APF measurements were obtained over a similar spectral range, but at a higher spectral resolving power, $R\sim$108,000. For each spectrum that was obtained, the region containing the iodine lines was divided into $\sim$700 chunks, each of $\sim2\,\AA$ width. Each chunk produces an independent measure of the wavelength, PSF, and Doppler shift. The final measured velocity is the weighted mean of the velocities of the individual chunks. All RVs have been corrected to the solar system barycenter, but are not tied to any absolute RV system. As such, they are ``relative'' RVs, with a zero point that is usually set simply to the mean of each set.

 The internal uncertainties quoted for all the radial velocity measurements in this paper reflect only one term in the overall error budget, and result from a host of systematic errors from characterizing and determining the point spread function, detector imperfections, optical aberrations, effects of undersampling the iodine lines, and other effects. Two additional major sources of error are photon statistics and stellar ``jitter''. The latter varies widely from star to star, and can be mitigated to some degree by selecting magnetically inactive older stars and by time-averaging over the star's unresolved low-degree surface $p$-modes. The median signal to noise (S/N) of our observations from Keck/HIRES in the iodine region used to calculate the Doppler radial velocity shift is 217. The APF telescope shows a similar S/N but with less variance owing to its state of the art precision. With a bright star such as HD~141399, the exposure time to attain this S/N is short enough that the star's $p$-mode oscillations are a real concern when determining its radial velocity. To avoid being dominated by $p$-modes, all observations in this paper have been binned on 2-hour timescales. The calculation of the binned velocity and associated timestamp take into account the internal uncertanty of each contributing spectrum and creates our observational data listed in Tables \ref{tab:rvdata_KECK} and \ref{tab:rvdata_APF}.

 \section{HD~141399 - Stellar Properties}
\begin{figure}
 \plotone{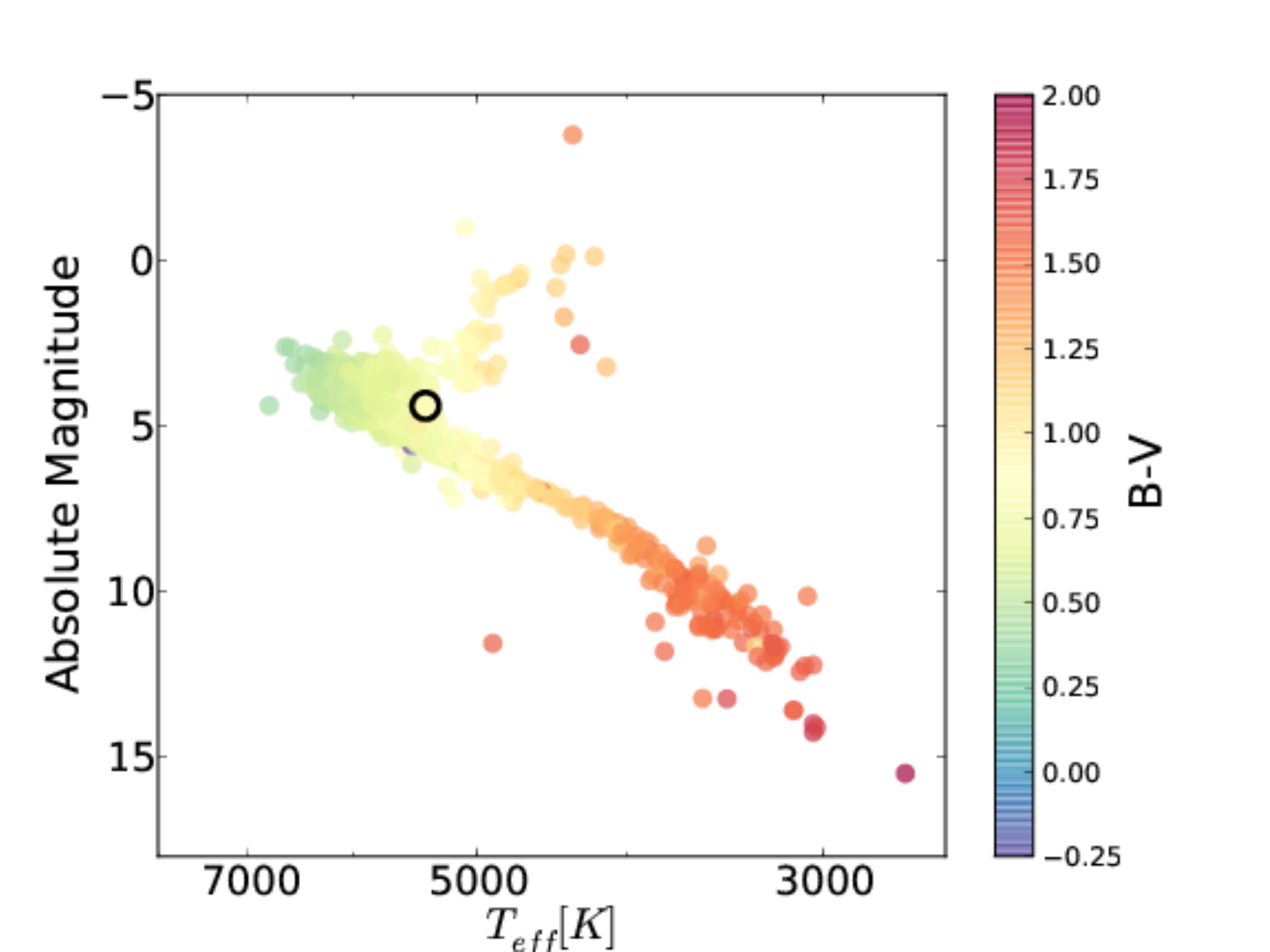}
 \caption{HR diagram with HD~141399's position indicated as a small open circle. Absolute magnitudes, $M$, are estimated from V band apparent magnitudes and Hipparcos distances using $M=V+5\log_{10}(d/10~{\rm pc})$. All 956 stars in our catalog of radial velocity measurements for which more than 20 Doppler measurements exist are shown, color-coded by their B-V values.}
 \label{fig:HRDiag}
 \end{figure}

\begin{figure}
 \plotone{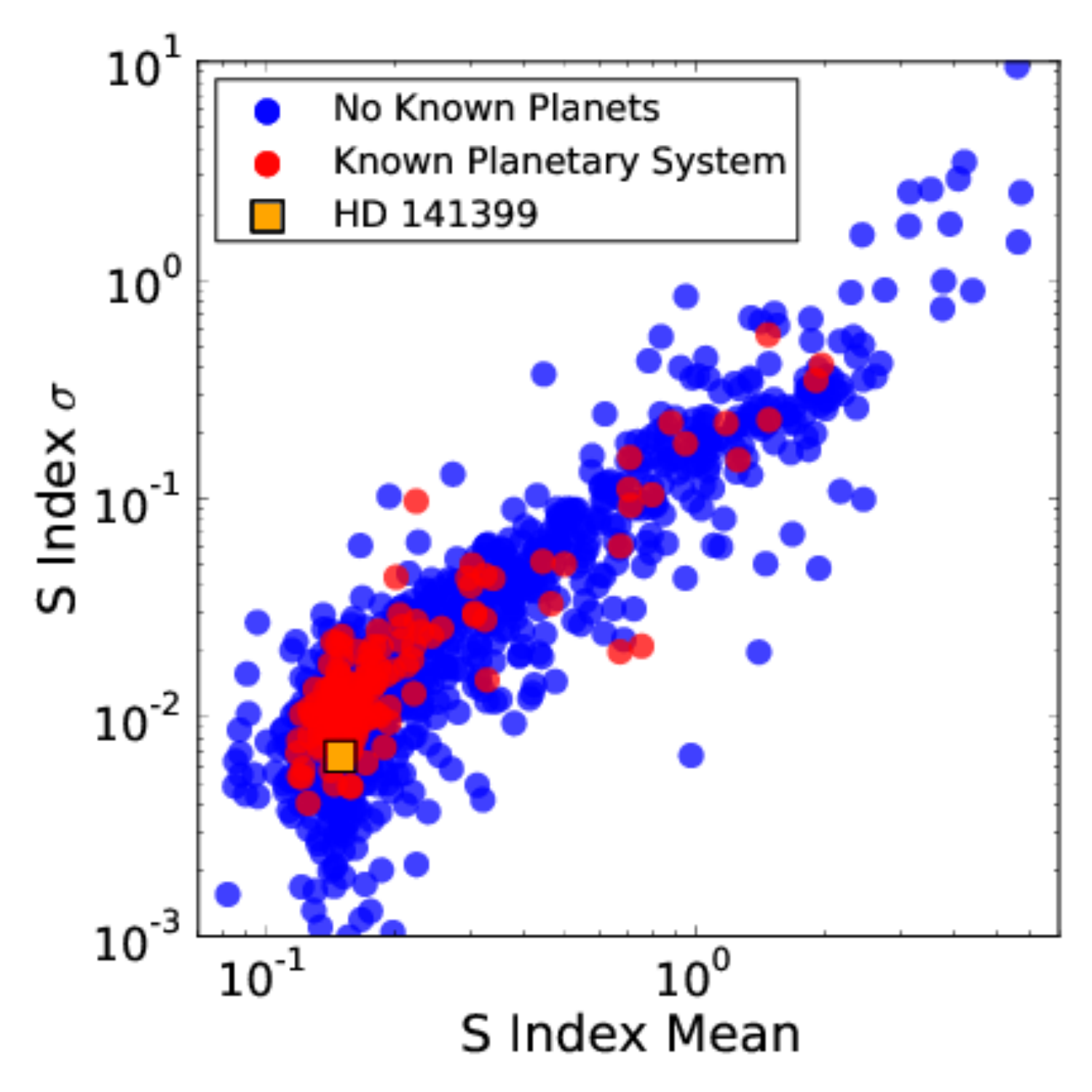}
 \caption{The average value of the S-Index against the standard deviation of the S-Index for all the stars in the Lick-Carnegie database. Planets with published planetary systems are colored red. HD~141399 is shown as a square near the quiet tail of the diagram.}
 \label{fig:Sindicies}
 \end{figure}

 HD 141399 (HIP 77301) is located at RA = +15:46:53.8 DEC = +46:59:10.5. As shown in Figure \ref{fig:HRDiag}, which is a color-magnitude diagram of all of the stars in the current Lick-Carnegie Keck database, the star lies just at the main sequence turnoff, with a ${\rm B-V}$ magnitude of $0.77 \pm 0.02$ \citep{VanLeeuwen07}. Following \citet{Torres10} we derive an effective temperature $T_{\rm eff}$ = $5360 \pm 53$~K for this star, which, when combined with the $d=36.17$ pc Hipparcos distance \citep{VanLeeuwen07}, and $V=7.2$ magnitude imply a stellar luminosity $L=1.59\,\pm 0.39\,L_{\odot}$ and a stellar radius $R=1.46\,\pm 0.15\,R_{\odot}$. Assuming ${\rm M}\sim {\rm L}^{1/3.5}$ gives us a stellar mass estimate of $1.14\,\pm 0.08\,M_{\odot}$. Using calibrations derived by \citet{Ammons06}, we estimate a metallicity, ${\rm Fe/H}=0.18 \pm 0.16$ for HD~141399, in keeping with the presence reported herein of several giant planets. Stars with detectable giant planets similar to the ones reported here have, on average, super-solar metallicities \citep{FischerValenti05}. HD~141399 has no known stellar companions, and is chromospherically quiet, with a Mt. Wilson S-index value $S_{\rm HK}=0.16$ \citep{Isaacson10} which implies an expected level of stellar jitter $\sigma_{\rm jitter} \le 2.25\, {\rm m\,s^{-1}}$. Indeed, as shown in Figure \ref{fig:Sindicies}, HD~141399's S-index places it among the locus of lowest activity stars within the current Keck survey list.

 \begin{deluxetable}{ccc}
 \tablecaption{KECK radial velocities for HD~141399
 \label{tab:rvdata_KECK}}
 \tablecolumns{3}
 \tablehead{{JD [UTC]}&{RV [m/s]}&{Uncertainty [m/s]}}
 \startdata
 2452833.85 & -58.53 & 1.29  \\
2452834.82 & -63.69 & 1.31  \\
2453179.91 & 30.70 & 1.15  \\
2453427.11 & -34.73 & 1.12  \\
2453807.07 & -35.18 & 1.38  \\
2454131.17 & 30.53 & 1.10  \\
2454549.04 & 57.35 & 1.49  \\
2454839.15 & -41.12 & 1.20  \\
2454865.14 & -53.61 & 1.24  \\
2454930.01 & -0.22 & 1.27  \\
2454955.93 & 19.04 & 1.55  \\
2454963.90 & 22.92 & 1.61  \\
2455014.81 & -4.82 & 1.12  \\
2455015.86 & -7.52 & 1.24  \\
2455041.84 & -62.07 & 1.12  \\
2455042.82 & -62.18 & 0.91  \\
2455043.76 & -63.51 & 1.05  \\
2455073.77 & -48.92 & 1.06  \\
2455106.73 & -17.28 & 1.24  \\
2455111.70 & -23.13 & 1.34  \\
2455133.69 & -8.57 & 1.27  \\
2455134.69 & -8.38 & 1.38  \\
2455135.69 & -5.75 & 1.25  \\
2455197.17 & 34.79 & 1.22  \\
2455198.15 & 30.91 & 1.28  \\
2455229.07 & -28.30 & 1.16  \\
2455232.08 & -32.45 & 1.29  \\
2455256.17 & -25.60 & 1.17  \\
2455257.00 & -25.64 & 1.21  \\
2455261.07 & -24.80 & 1.38  \\
2455284.97 & -23.37 & 1.32  \\
2455286.00 & -18.45 & 1.23  \\
2455290.04 & -20.90 & 1.32  \\
2455312.91 & -15.48 & 1.21  \\
2455313.87 & -10.42 & 1.07  \\
2455319.91 & -4.57 & 1.30  \\
2455320.88 & -6.76 & 1.25  \\
2455321.89 & 1.12 & 1.25  \\
2455342.83 & 40.69 & 1.10  \\
2455344.84 & 44.13 & 1.15  \\
2455350.79 & 59.22 & 1.13  \\
2455351.87 & 66.16 & 1.12  \\
2455372.80 & 78.43 & 1.17  \\
2455377.81 & 74.54 & 1.09  \\
2455380.81 & 72.17 & 1.01  \\
2455396.86 & 42.95 & 1.13  \\
2455399.95 & 37.39 & 1.18  \\
2455400.78 & 35.93 & 1.06  \\
2455401.77 & 33.07 & 1.02  \\
2455402.83 & 27.90 & 1.13  \\
2455403.83 & 31.25 & 1.16  \\
2455404.78 & 29.19 & 1.00  \\
2455405.77 & 27.97 & 1.00  \\
2455410.79 & 20.87 & 1.05  \\
2455669.98 & -15.70 & 1.06  \\
2455719.99 & 9.04 & 1.17  \\
2455749.93 & 61.14 & 0.94  \\
2455824.81 & 12.16 & 1.14  \\
2455825.78 & 10.18 & 0.98  \\
2455839.71 & 0.00 & 1.07  \\
2455840.76 & 0.94 & 1.22  \\
2455971.10 & 13.23 & 1.13  \\
2455972.11 & 11.52 & 1.09  \\
2456025.01 & 4.30 & 1.14  \\
2456027.00 & -0.40 & 1.21  \\
2456116.96 & -21.80 & 1.09  \\
2456117.92 & -23.82 & 0.95  \\
2456168.77 & -3.17 & 1.08  \\
2456169.79 & -4.53 & 1.06  \\
2456291.16 & -52.97 & 1.09  \\
2456329.17 & 4.55 & 1.19  \\
2456330.15 & -1.20 & 1.02  \\
2456432.86 & 0.74 & 1.12  \\
2456433.94 & 1.71 & 1.09  \\
2456548.72 & 20.25 & 1.07  \\
2456549.74 & 21.67 & 0.97  \\
2456551.76 & 23.66 & 0.94  \\

 \enddata
 \end{deluxetable}
 \begin{deluxetable}{ccc}
 \tablecaption{APF radial velocities for HD~141399
 \label{tab:rvdata_APF}}
 \tablecolumns{3}
 \tablehead{{JD [UTC]}&{RV [m/s]}&{Uncertainty [m/s]}}
 \startdata
 2456482.80 & -29.02 & 0.87  \\
2456488.70 & -19.66 & 0.80  \\
2456490.71 & -14.51 & 0.96  \\
2456492.71 & -16.93 & 0.77  \\
2456494.71 & -18.79 & 0.86  \\
2456510.74 & 0.00 & 0.79  \\
2456521.78 & 3.43 & 0.79  \\
2456532.73 & 8.55 & 0.86  \\
2456533.72 & 6.76 & 0.84  \\
2456551.69 & 23.96 & 1.11  \\
2456553.67 & 23.74 & 0.96  \\
2456622.05 & 30.47 & 0.71  \\
2456662.09 & -27.64 & 0.98  \\
2456677.90 & -15.84 & 0.78  \\

 \enddata
 \end{deluxetable}

 \begin{figure}
 \plotone{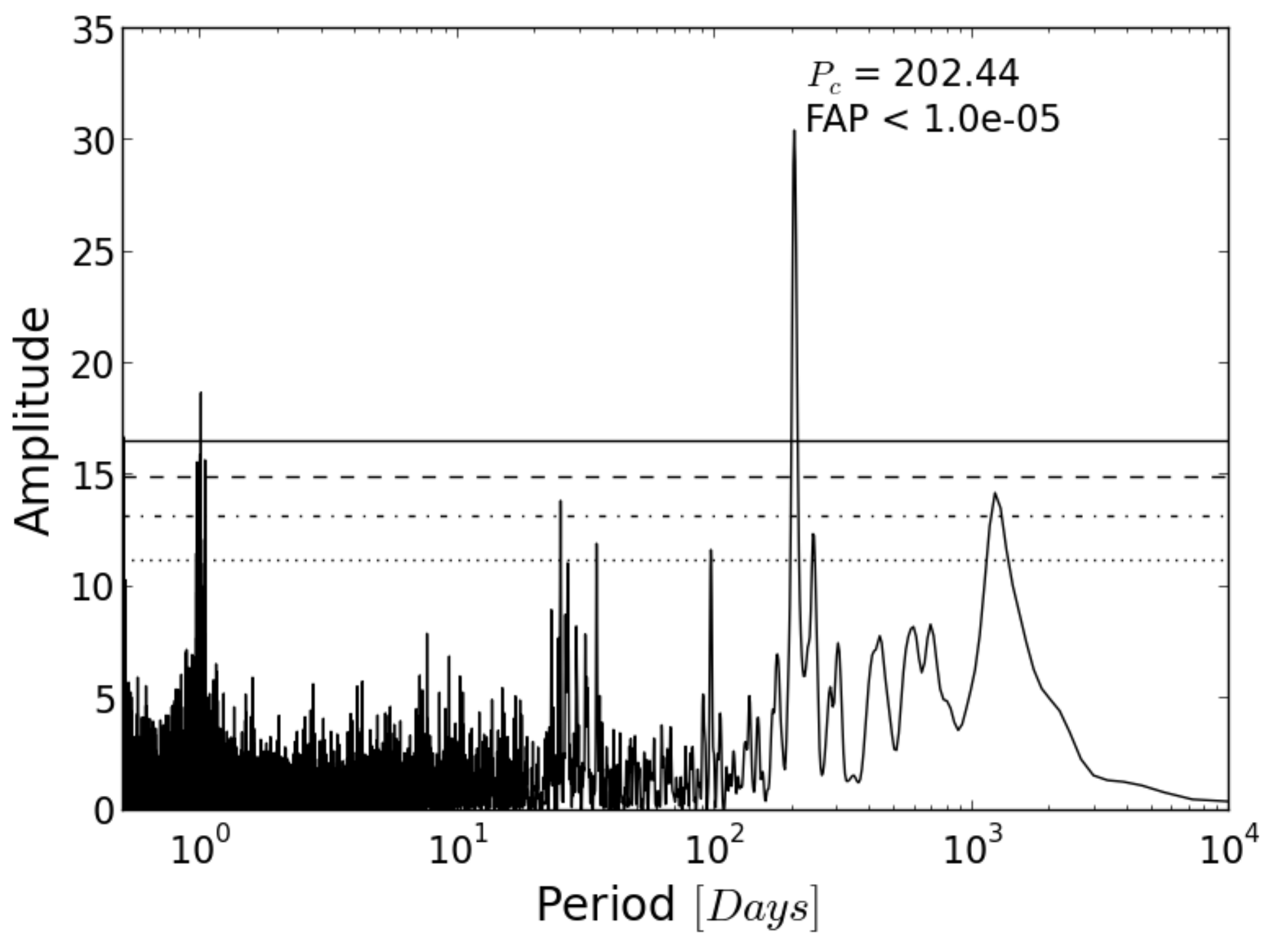}
 \plotone{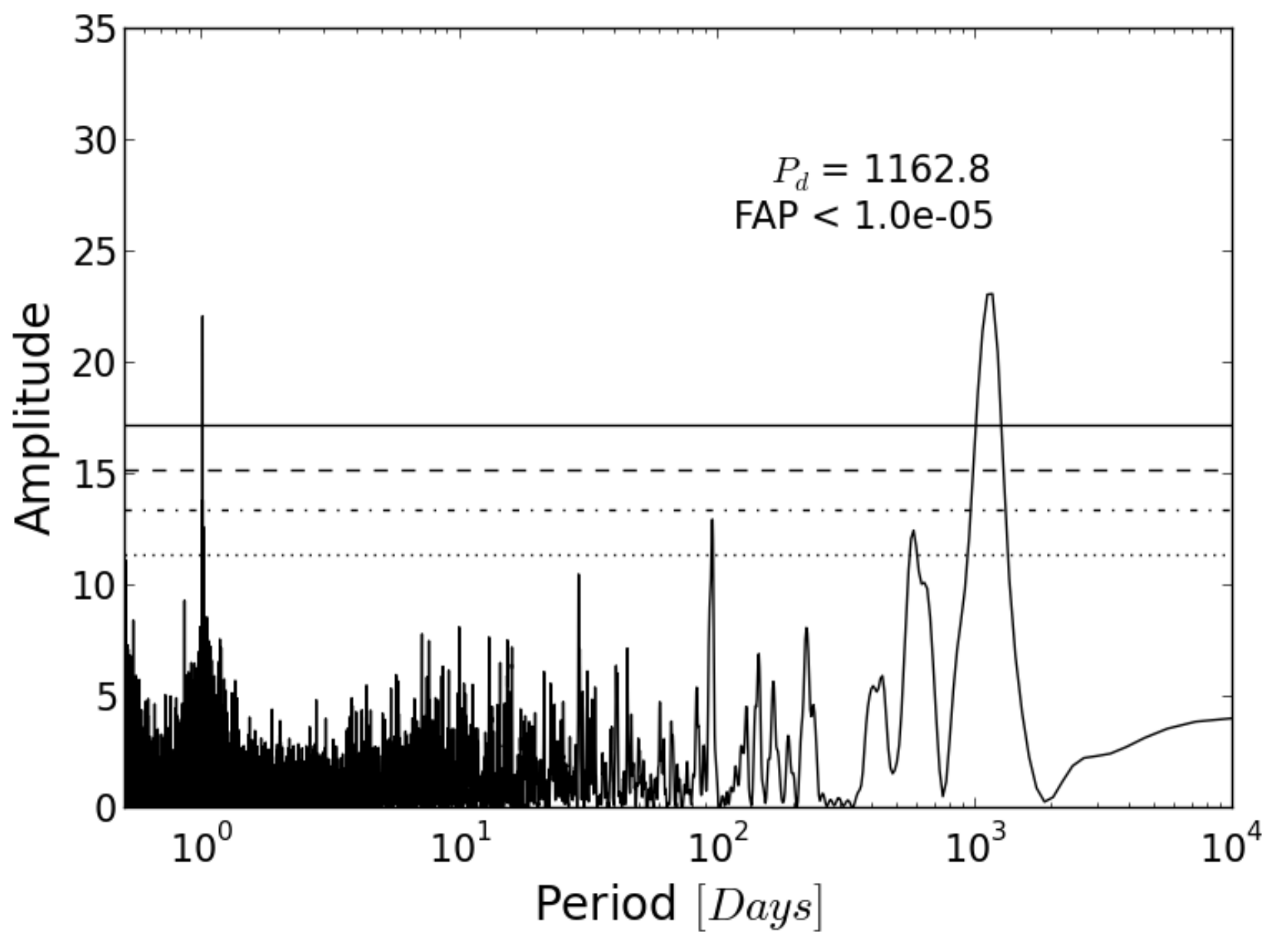}
 \plotone{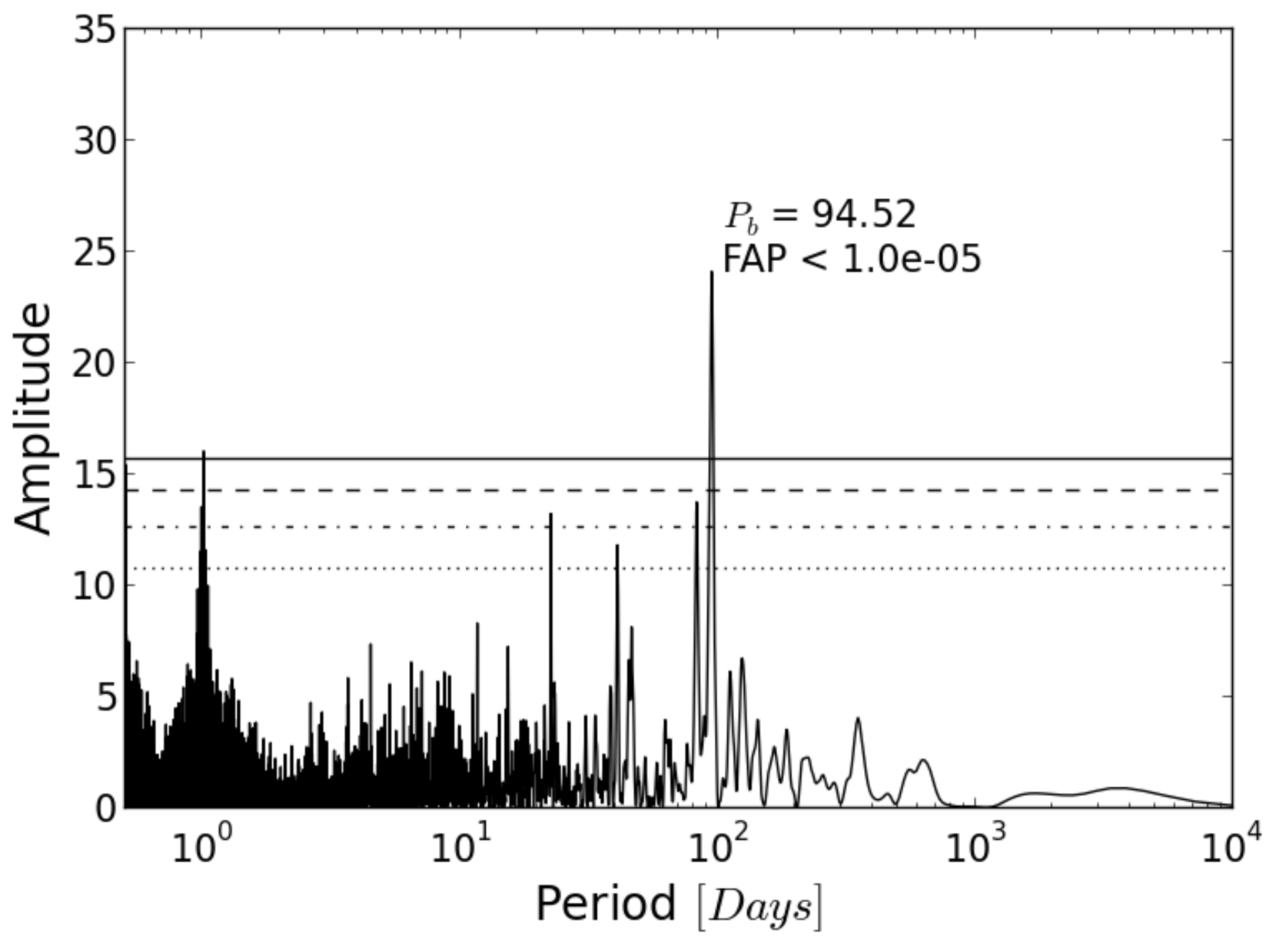}
 \plotone{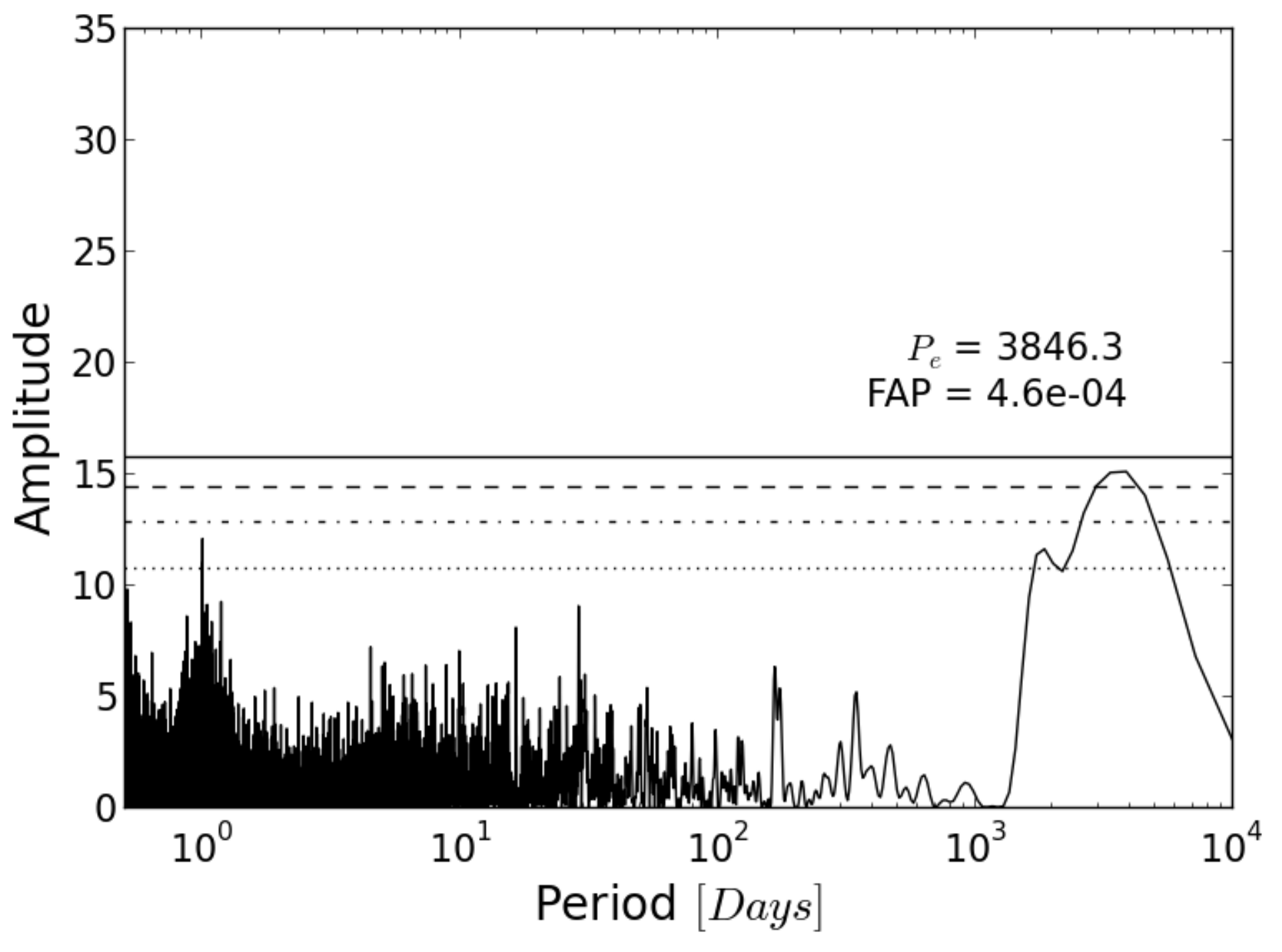}
 \caption{Lomb-Scargle periodograms for {\it top panel:} radial velocity measurements of HD 141399 from the Keck and APF telescopes, {\it second panel from top} residual velocities with planet c removed, {\it second panel from bottom} residual velocities with planets c and d removed, {\it bottom panel} residual velociteis with planets b, c, and d removed. The horizontal lines from top to bottom represent false alarm probabilities of 0.01\%, 0.1\%, 1.0\% and 10.0\% respectively.}\label{fig:periodogram}
 \end{figure}

 \begin{figure}
 \plotone{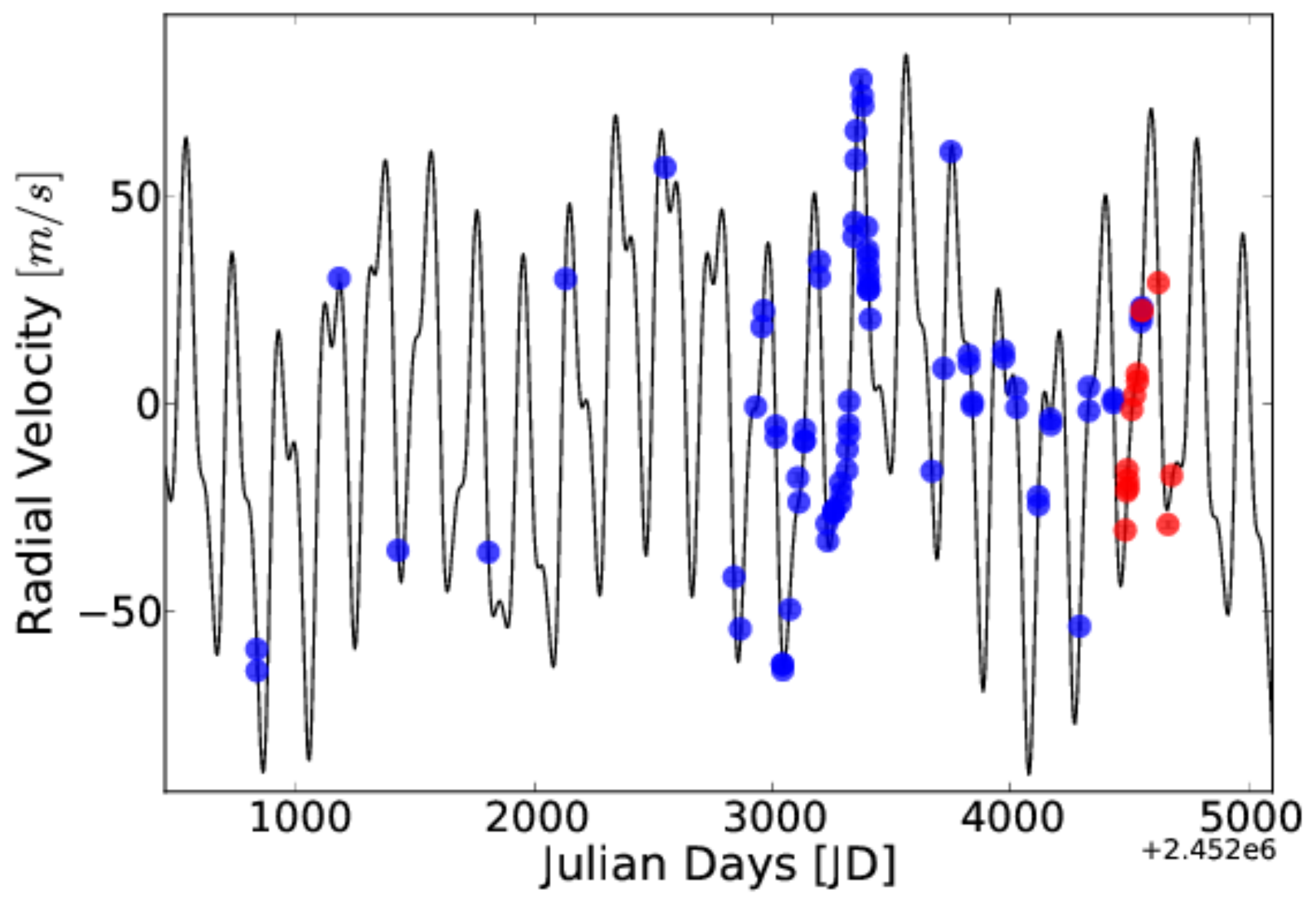}
 \plotone{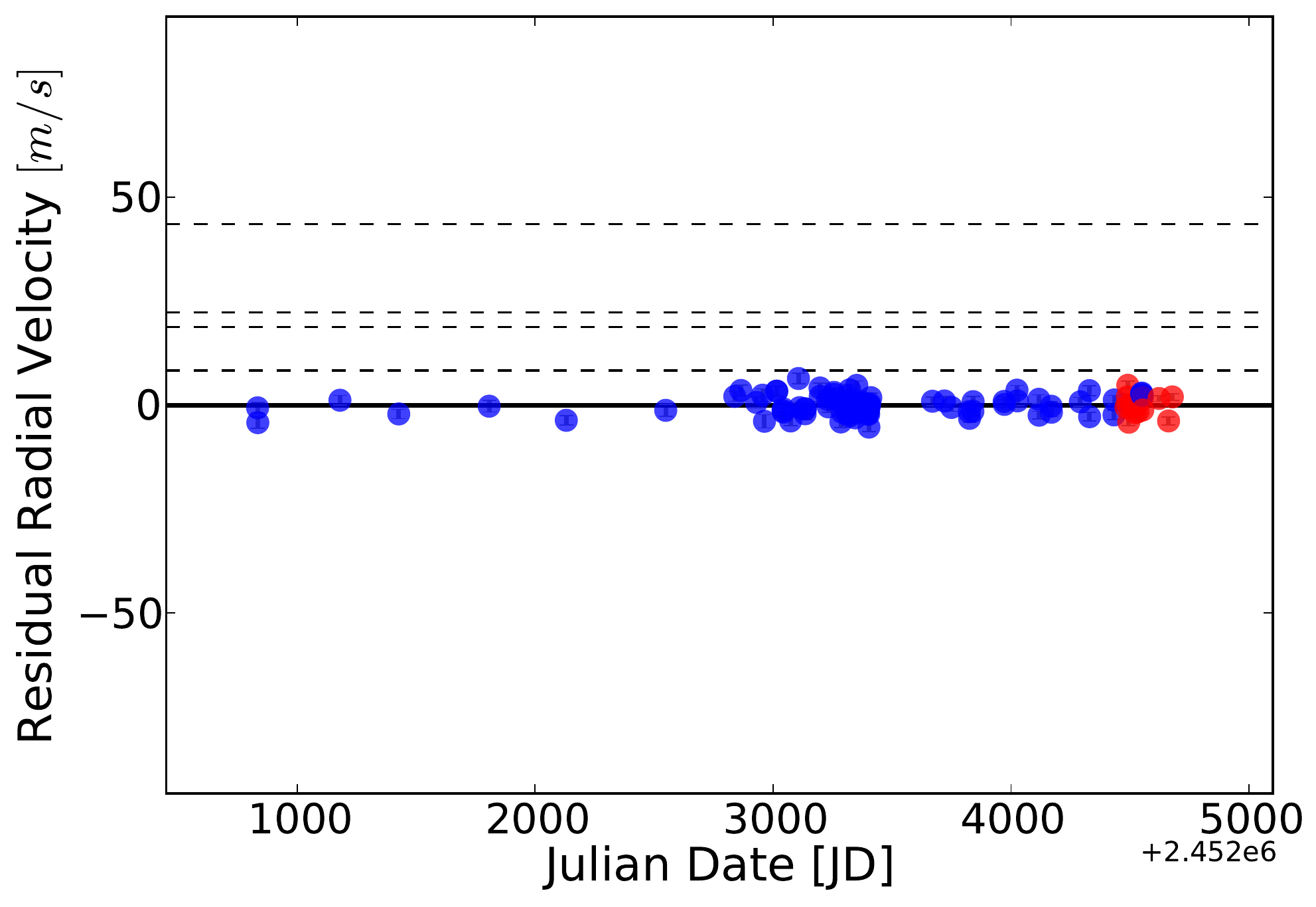}
 \plotone{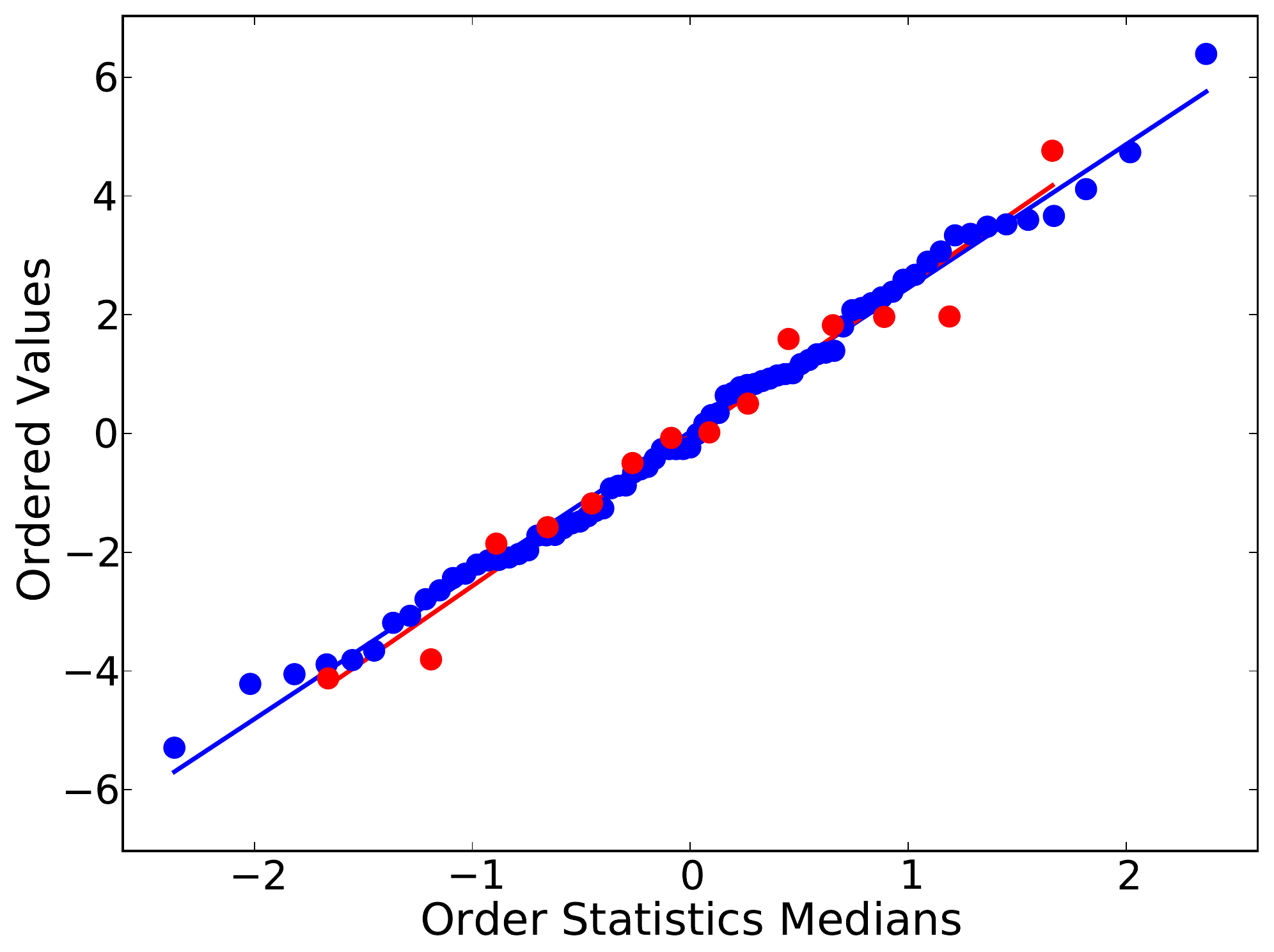}
 \caption{{\it top panel} Best-fit self-consistent integrated 4-planet model from Table \ref{tab:fit}, integrated and compared to the RV measurments for HD~141399. RVs obtained with Keck/HIRES are blue, RVs from the APF/Levy are shown in red. {\it middle panel} Velocity residuals to the best fit 4-planet model of the system. The four dashed lines indicate from top to bottom ${\rm K}_c =43.08 {\,\rm m\,s^{-1}}$, ${\rm K}_d =21.95 {\,\rm m\,s^{-1}}$, ${\rm K}_b =18.6 {\,\rm m\,s^{-1}}$, ${\rm K}_e =9.42 {\,\rm m\,s^{-1}}$, the radial velocity half-amplitudes of the detected planets in the system. {\it bottom panel} Quantile-Quantile plot for the velocity residuals, indicating the degree to which the errors conform to a gaussian distribution.}
 \label{fig:fitPlot}
 \end{figure}

 \begin{deluxetable}{lllr}
 \tablecaption{Self-consistent 4-planet model for the HD~141399 System\label{tab:fit}}
 \tablecolumns{4}
 \tablehead{{}&{}&{Best fit}&{Errors}}
 \startdata
 Period (d)&b&94.35&(0.059) \\
&c&202.08&(0.099) \\
&d&1070.35&(8.178) \\
&e&3717.35&(555.081) \\
& & & \\
RV Half-Amplitude ($m\,s^{-1}$)&b&18.8&(0.551) \\
&c&43.51&(0.591) \\
&d&22.28&(0.63) \\
&e&8.34&(1.239) \\
& & & \\
Mean Anomaly (deg)&b&224.63&(54.09) \\
&c&303.75&(15.165) \\
&d&273.89&(39.812) \\
&e&153.93&(23.889) \\
& & & \\
Eccentricity&b&0.04&(0.03) \\
&c&0.05&(0.013) \\
&d&0.06&(0.029) \\
&e&0.0&(Fixed) \\
& & & \\
Longitude of Periastron (deg)&b&191.37&(55.088) \\
&c&214.74&(14.457) \\
&d&249.16&(38.966) \\
&e&0.0&(Fixed) \\
& & & \\
Time of Periastron (JD)&b&2452774.98&(15.371) \\
&c&2452663.34&(8.537) \\
&d&2452019.53&(119.538) \\
&e&2451244.36&(555.624) \\
& & & \\
Semi-Major Axis (AU)&b&0.4225&(0.00018) \\
&c&0.7023&(0.00023) \\
&d&2.1348&(0.01086) \\
&e&4.8968&(0.46122) \\
& & & \\
Mass ($M_{\rm Jup}$)&b&0.46&(0.025) \\
&c&1.36&(0.067) \\
&d&1.22&(0.067) \\
&e&0.69&(0.164) \\

 &&&\\
 First Observation Epoch (JD)&& 2452833.85&\\

 Velocity Offset (KECK) && 0.61 $\,{\rm m\,s^{-1}}$ & (1.7) \\
 Velocity Offset (APF) && 1.48 $\,{\rm m\,s^{-1}}$ & (1.89) \\
 $\chi^2$ && 5.81&\\
 RMS && 2.36 $\,{\rm m\,s^{-1}}$ &\\
 Jitter (KECK) && 2.35 $\,{\rm m\,s^{-1}}$ & (0.281)\\
 Jitter (APF) && 2.59 $\,{\rm m\,s^{-1}}$ & (0.729)\\
 \enddata
 \tablecomments{All elements are defined at epoch JD = 2452833.85. Uncertainties are reported in parentheses.}
 \end{deluxetable}

 \begin{figure}
 \plotone{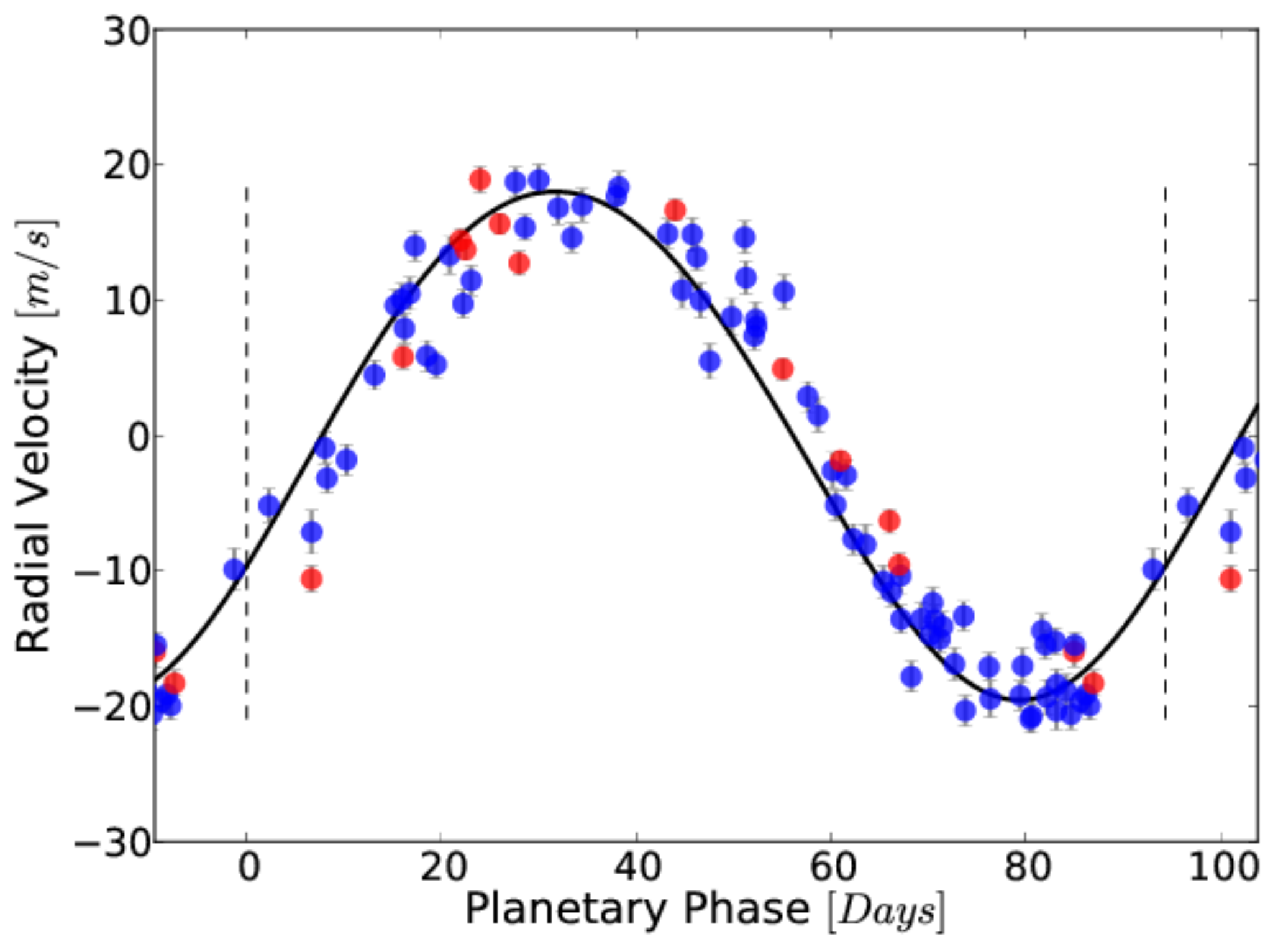}
 \plotone{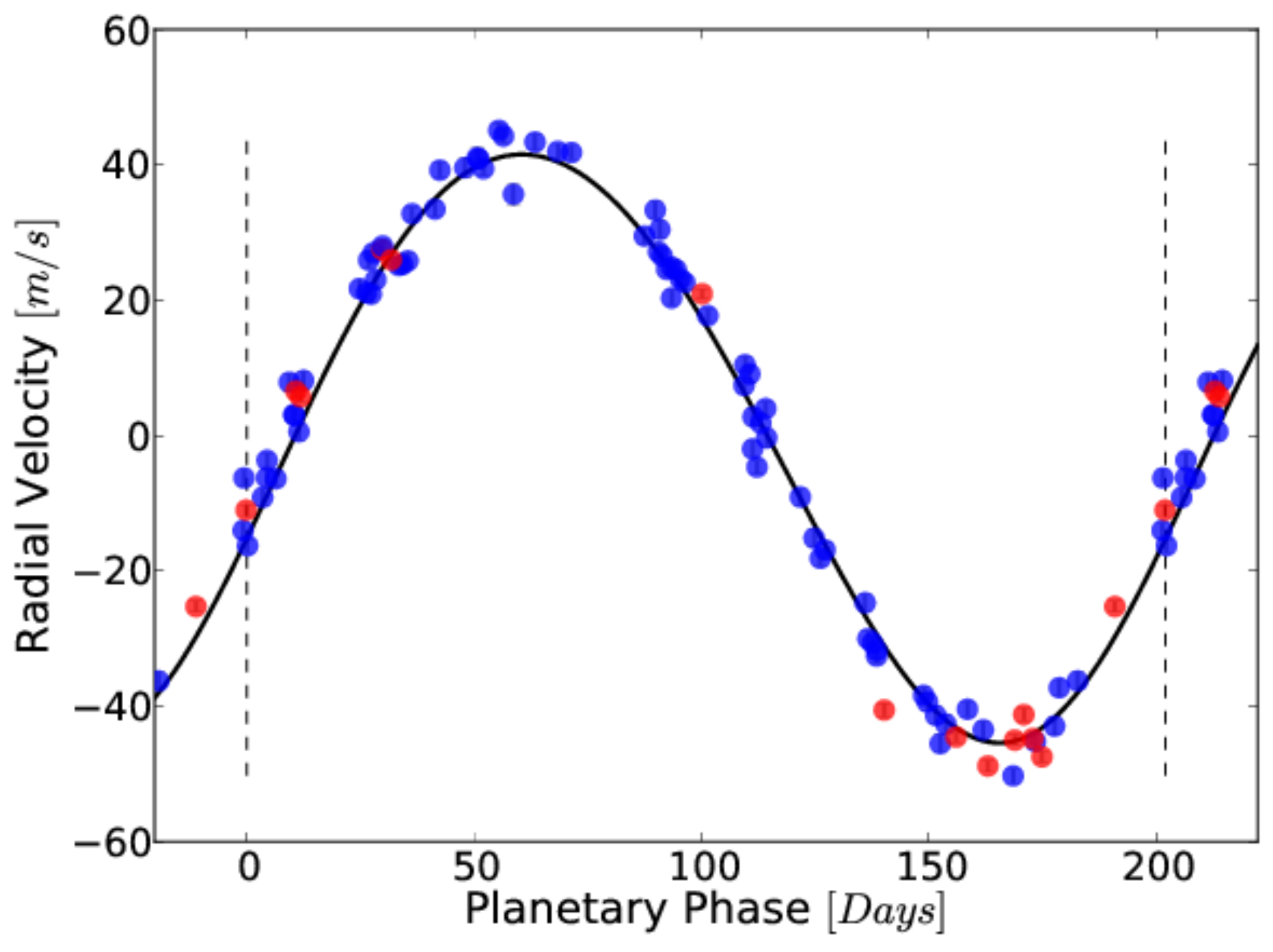}
 \plotone{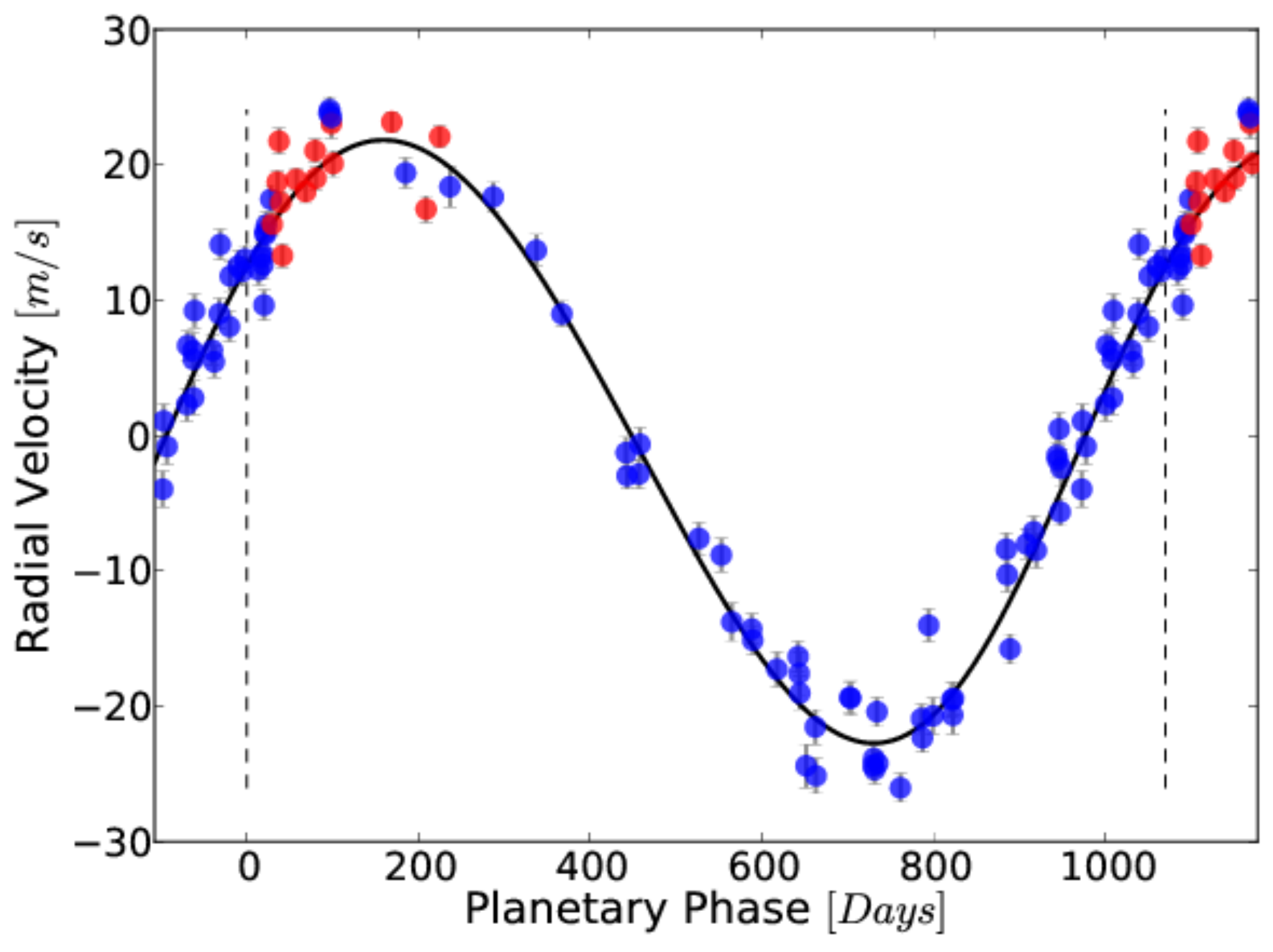}
 \plotone{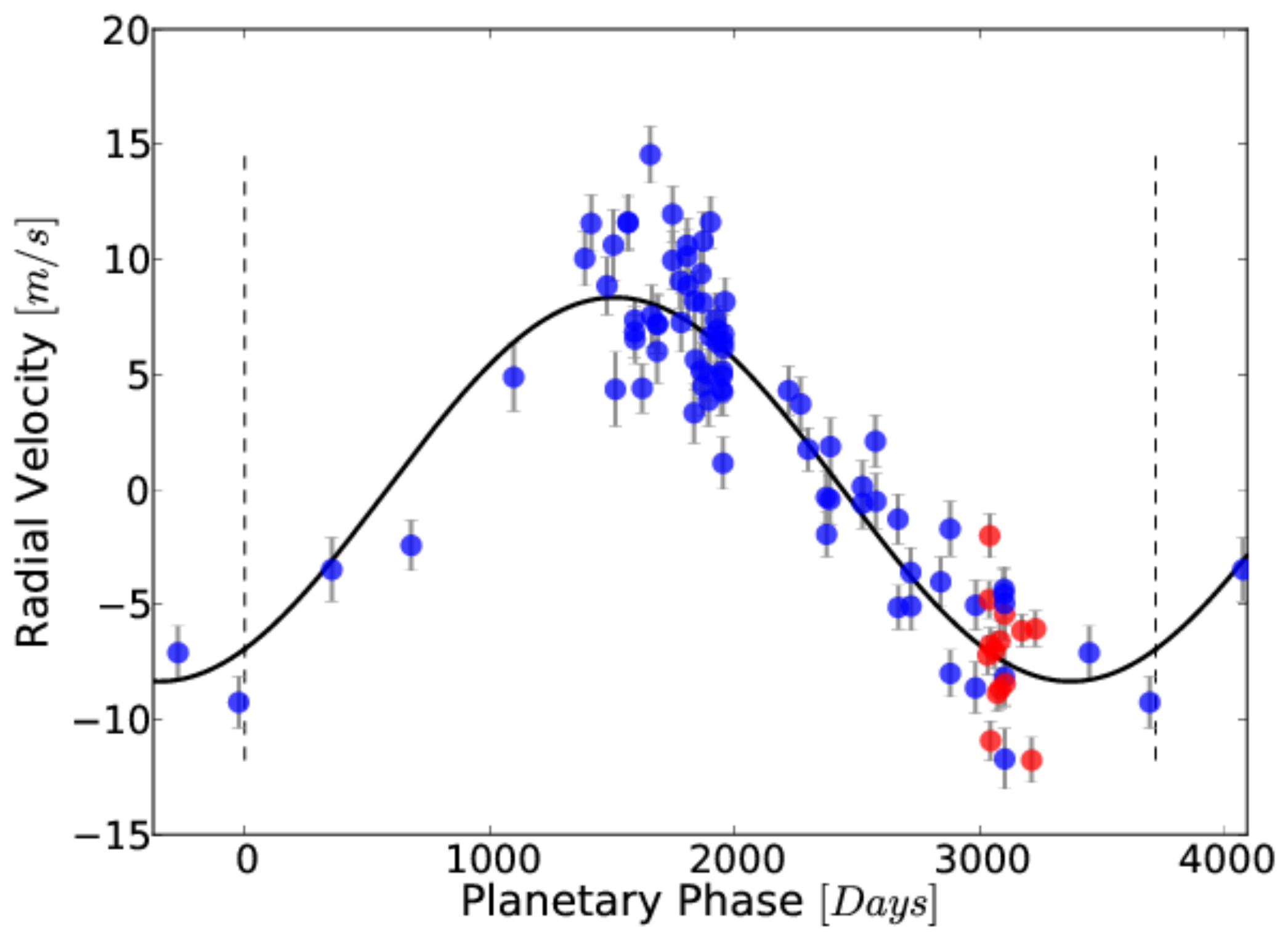}
 \caption{Phased RV curves for planets b, c, d, and e. The error estimate for each RV data point is also plotted, but may not be visible due to the scaling of the individual curves. The vetical dashed lines denote the extent of unique data.}
 \label{fig:phasedCurves}
 \end{figure}

 \begin{figure}
 \plotone{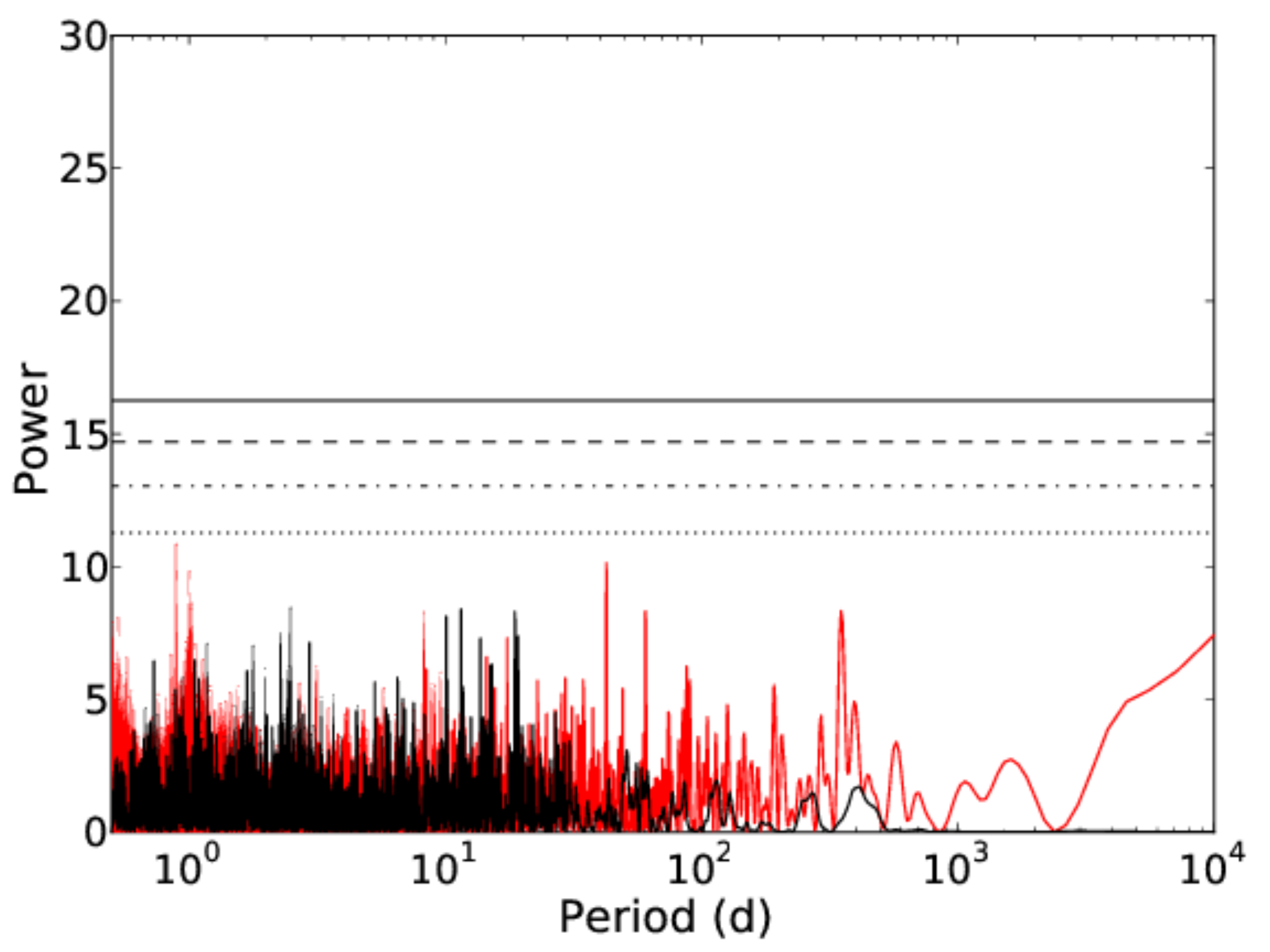}
 \caption{Lomb-Scargle periodogram of the radial velocity residuals to the fit given in Table \ref{tab:fit} plotted in black, and the Lomb-Scargle periodogram of the Mt. Wilson S-Index values plotted behind in red.}
 \label{fig:residualPer}
 \end{figure}

 \begin{figure*}
 \centering
 \includegraphics[height=14.0cm,angle=0]{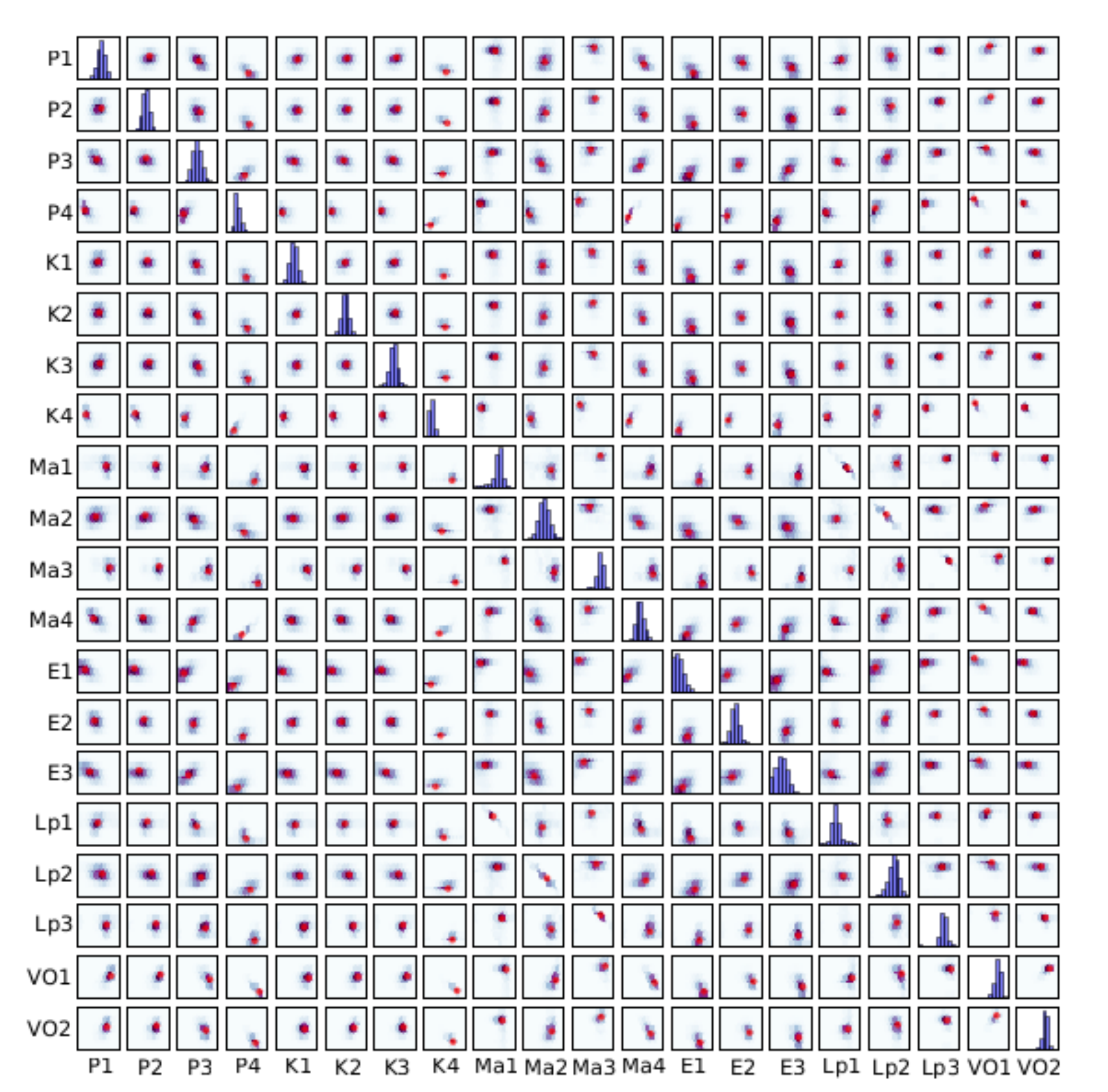}
 \caption{Smooth scatter plot of parameter error correlations for our Markov chain. In each case, the best fit model is indicated with a small red dot, and the density of models within the converged portion of the chain is shown as a blue toned probability distribution function.}
 \label{fig:mcmcErrorLarge}
 \end{figure*}

 \begin{figure}
 \plotone{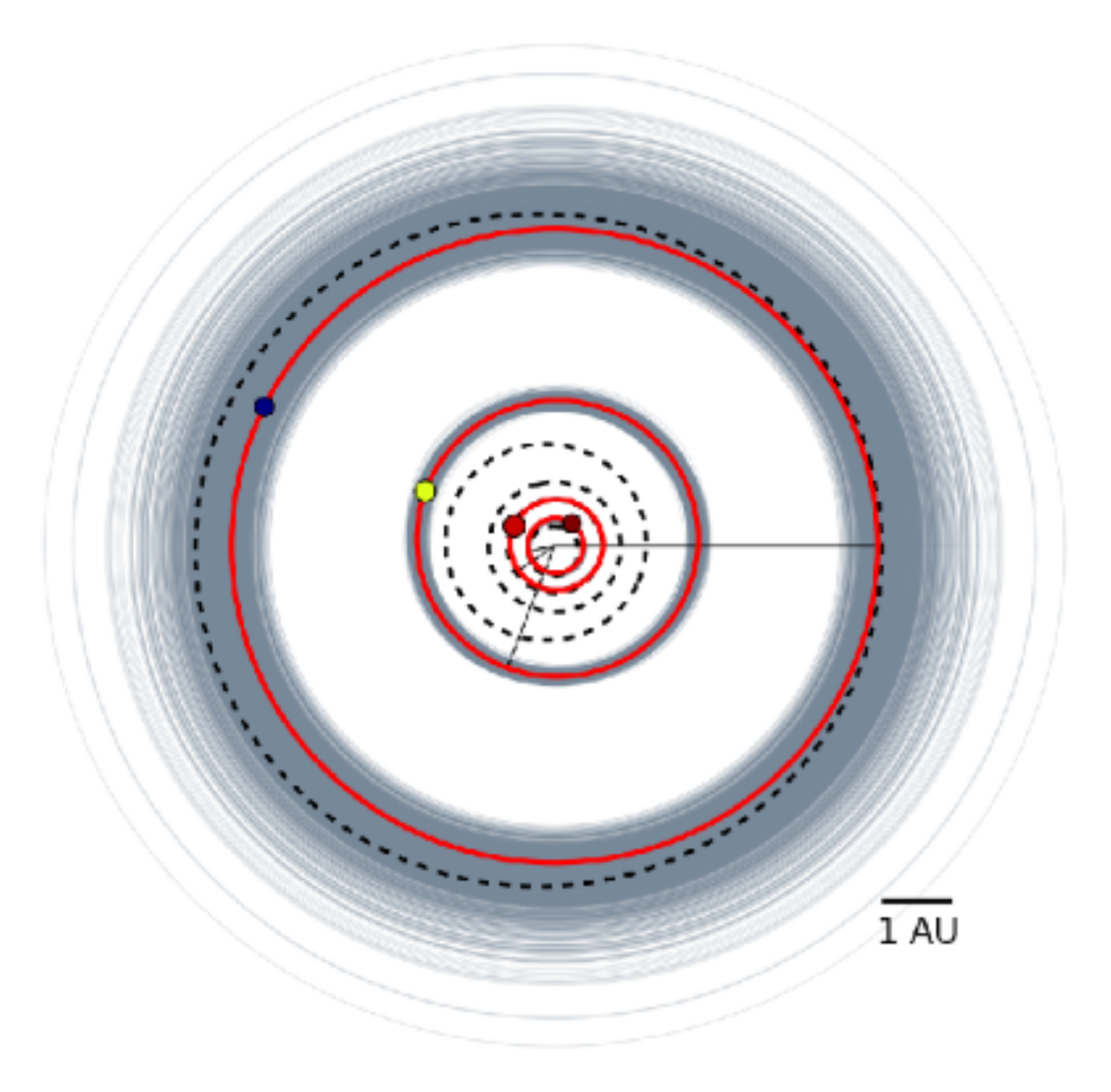}
 \caption{The orbits of the proposed planetary system around HD~141399. The points correspond to the location of the planets at the initial observation epoch 2452833.85. The lines from the origin correspond to each planet's perihelion. The light lines are 1000 orbits of the planets drawn from the converged segment of the Markov Chain. The dashed lines are the orbits of the solar system planets with Mercury, Venus, Earth, Mars, and Jupiter all shown here.}
 \label{fig:orbitDiag}
 \end{figure}

 \section{Best-fit solution}
 The combined radial velocity data sets show a root-mean-square (RMS) scatter of 32.8 ${\rm m\,s}^{-1}$ about the mean velocity. This includes a ``telescope'' offset of $\sigma_{\rm tel}\sim -0.87\, {\rm m\,s}^{-1}$, between the Keck and APF velocity zero points. (The exact value of $\sigma_{\rm tel}$ is allowed to vary as a free parameter and emerges as a measured quantity in Table \ref{tab:fit}). HD 141399's quiet chromosphere implies a low level of expected stellar jitter, $\sigma_{\rm jitt}<2\,{\rm m\,s}^{-1}$, which is consistent with a planetary explanation for the observed dispersion in radial velocities.

 A Lomb-Scargle periodogram of the 91 velocity measurements of HD~141399 is shown in the top panel of Figure \ref{fig:periodogram}. False alarm probabilities are calculated with the bootstrap method, as described in \citet{Efron79}, iterating 100,000 times for a minimum probability of 1e-5 as seen on the $P_{\rm b}$ , $P_{\rm c}$ , and $P_{\rm d}$ peaks in the top panels of Figure \ref{fig:periodogram}. Several periodicities having vanishingly small false alarm probabilities are observed, with the strongest lying at $P_{\rm c}=201.88$ days. This signal in the data is modeled as a $M_{\rm c}\sin(i)=1.36\,M_J$ planet with an orbital eccentricity, $e_c=0.05$. The resulting periodogram of the residuals is shown in the second panel from the top of Figure \ref{fig:periodogram}, and indicates the presence of significant power at periods of both $P_{\rm d}=1163$ and $P_{\rm b}=94.52$ days. These two signals are modeled with planets of masses $M_{\rm b}\sin(i)=0.46\,M_{\rm J}$ and $M_{\rm d}\sin(i)=1.22\,M_{\rm J}$. The residuals periodogram to the three-planet fit suggests the presence of a fourth planet in the system, with properties that are quite reminiscent of Jupiter, $P_{\rm e}\sim 4000$ days, and $M_{\rm e}\sin(i)\sim 1.0\,M_{\rm J}$.

 Using Levenberg-Marquardt optimization, we generated a 4-planet best-fit Keplerian model for the system. We note that the proximity of the 94-day planet and the 201-day planet to a 2:1 mean motion resonance generates a modest dynamical interaction over the period covered by the radial velocity measurements. For our preferred model of the system, we therefore derived a Newtonian fit in which planet-planet interactions (dominated by steady orbital precession, ${\dot {\omega}}_b$, of the inner planet) are taken explicitly into account. This self-consistent fit (which additionally assumes $i=90^{\circ}$ and $\Omega=0^{\circ}$ for all four planets) is listed in Table \ref{tab:fit}, and the stellar reflex velocity arising from our 4-planet orbital model is compared to the radial velocity time series in the top panel of Figure \ref{fig:fitPlot}. Phased RV curves for each of the four planets in Table \ref{tab:fit} are shown in Figure \ref{fig:phasedCurves}. A power spectrum of the residuals to our four-planet fit is shown in Figure \ref{fig:residualPer} and indicates no significant periodicities. Also shown in this figure is a periodogram of the Mt. Wilson S-Index. None of the peaks in the periodogram of S-Index values coincide with peaks that we are interpreting to arise from planets. Although there is a long term trend present in the S-Index periodogram, it does not have a well defined peak like that of planet e, shown in Figure \ref{fig:periodogram}.

 In addition to stellar activity, another potential source of false positive detections stems from aliases that arise in response to the observational sampling. Because our observations are constrained by when the star is visible in the night sky, and because Keck Telescope time is allocated to Doppler surveys primarily when the Moon is up, we expect aliases at periods of 1 solar day, 1 sidereal day, 1 synodic month and 1 sidereal year. Examining the window function we
 do see peaks resulting at these periods, but careful analysis of the periodogram for our radial velocity observations, as described in \citet{Dawson10}, shows no evidence of strong signals occurring at the locations necessary for any of our four planetary signals to be potential aliases.

 The reduced chi-squared statistic for our dynamically self-consistent 4-planet fit is $\chi_{red}^2 =$ 5.81 and results in a fit with an RMS of 2.36 ${\rm m\,s^{-1}}$ and estimated stellar jitter of $\sigma_{\rm jitter}$ = 2.07 ${\rm m\,s^{-1}}$ -- the estimate of the stellar jitter that is required to bring the reduced chi-squared statistic of the fit down to unity. Thus, if the true stellar jitter is of order 2.0 ${\rm m\,s^{-1}}$, then our four-planet fit adequately explains the excess variance in the radial velocity time series.

 In order to compute parameter uncertainties for our orbital fit, we implement a Markov Chain Monte Carlo algorithm \citep{Ford05, Ford06, Balan09, Meschiari09, Gregory11}. The MCMC algorithm returns a chain of state vectors, ${\bf k}_{i}$ (a set of coupled orbital elements, e.g. period, mass, etc. and the two velocity offset parameters). The goal of the Markov Chain calculation is to generate an equilibrium distribution proportional to $\exp[\chi^{2}({\bf k})]$. We adopt non-informative priors on all parameters (and uniform in the log for masses and periods). The resulting error correlations are shown in Figure \ref{fig:mcmcErrorLarge}, and a set of 1000 states drawn randomly from the converged chain are shown in an orbital diagram (Figure \ref{fig:orbitDiag}), which shows the orbits of the solar system planets for comparison.

 The error correlation diagram indicates that all parameters are well determined, save the usual degeneracies between mean anomaly and the longitude of periastron for the low-eccentricity orbits. Note that we have fixed the eccentricity of the outer planet to zero; it is not a parameter in the fit. If the eccentricity of the outer planet is allowed to float, one finds little change in the mass, but the best-fit model favors a Jupiter-mass planet on a very long period ($P > 40,000$ d), high eccentricity ($e > 0.8$), comet-like orbit, in which the planet makes a periastron passage that is centered on the time base line of our observations. We view such a configuration to be unlikely, and thus set $e_{\rm e} =0$ pending acquisition of more RV data. With $e_{\rm e}=0$, the distribution of the residuals relative to the best-fit model shows no evident pathologies. Indeed, a quantile-quantile plot (shown in the bottom panel of Figure \ref{fig:fitPlot}) indicates that the distribution of residuals is very well described by a normal distribution. As a consequence of the values $K/{\rm RMS} > 1$ for all the planets, the system appears to be well characterized, with the one exception being the orbital eccentricity of the outer planet.

 With an apparent V magnitude of 7.2, HD~141399 lies just below the threshold of naked-eye visibility and is brighter than all known hosts of transiting extrasolar planets other than 55 Cancri, which has V=6, and whose innermost $P=0.734$d, $M=8.4\,M_{\oplus}$ planet ``e'' has been observed in transit \citep{Winn11, Demory11}. As a consequence, transits by one or more of HD~141399's planetary companions, if they do occur, would be of substantial scientific value. The atmospheric chemistry of planets b and c is likely to be dynamic and complex, and so transmission spectroscopy with HST (or better, with JWST) would permit insight into the conditions governing planets that are bracketed in temperature by the well-studied hot Jupiters at one extreme and the even better studied Jovian planets of our own solar system at the other. The equilibrium temperature, $$T_{\rm eq}=({R_{\star}^{1/2} T_{\star} )/({(2a)^{1/2}(1-e^{2})^{1/8}}}),$$ for HD141399~b is 500K, and $T_{\rm eq}=390$K for planet c. The optical photospheres of both planets are potentially dominated by water clouds \citep{Marley07}, and it is possible that non-equilibrium chemistry and exotic compounds might influence both the atmospheric conditions and planetary meteorology.

 The \textit{a-priori} geometric probability, $\cal P$ that HD~141399~b can be observed in transit is ${\cal P}_{\rm b}=1.3$\%, and the odds that HD~141399~c transits are a slim 0.8\%. Adopting a super-solar metallicity for the star implies that all four planets in the system may have heavy element cores with $M_{\rm core}>30~M_{\oplus}$. The structural models of \citep{Bodenheimer03} predict radii of $R_{\rm b}=0.94\,R_{\rm J}$ and $R_{\rm c}=1.03\,R_{\rm J}$, implying transit depths, $d$, slightly larger than $d=0.5$\% for both planets.

 Planets such as HD~141399~b and c provide an opportunity to democratize access to cutting-edge research in exoplanetary science. Nearly all of the highly-cited ground-based discoveries of transiting extrasolar planets have been made with small telescopes of aperture $d<1$ m. Amateur observers, furthermore, were co-discoverers of the important transits of HD 17156b \citep{Barbieri07}, and HD 80606b \citep{Garcia-Melendo09}. The inner HD~141399 planets provide a good opportunity for such a campaign \citep[see e.g.,][]{Kane09}. Given the planetary model of Table 3, we find transit ephemerides for the two planets of $T_{\rm tran\, b}= {\rm JD}\, 2456615.83 + {\rm N}(94.4)$ and $T_{\rm tran\, c}={\rm JD}\,2456637.92 + {\rm N}(201.8)$. Here ${\rm N}(Period)$ is simply the period times any integer value. The high declination is well suited to observers in the Northern Hemisphere.

 \section{Dynamical Stability}

 A configuration of four giant planets in relatively close proximity raises questions regarding the dynamical stability and evolution of the system that can be easily addressed with numerical integrations. As discussed in the foregoing section, the relatively high values of $K/{\rm RMS}$ for HD~141399's inner three giant planets allow the orbital eccentricities to be determined fairly accurately, and with $e \lesssim 0.1$, for all three, the configuration is reminiscent (from the standpoint of dynamical stability) of the Jovian planets in our own solar system. While the lack of full orbital phase coverage for the observations of the outermost candidate planet, ``e'', render its eccentricity uncertain, its RV signature is consistent with a circular orbit.

 The orbital parameters listed in Table \ref{tab:fit} are osculating Jacobi elements referenced to JD 2452833.85, the epoch of the first Doppler velocity measurement of the star obtained with the Keck telescope. The substantial masses of the three planets, in conjunction with the relatively large number of orbits over which the inner planets have been observed and the proximity to the 2:1 mean-motion resonance for planets b and c, produce a tangible degree of interaction between the planets during the decade that they have been monitored. For example, if the best-fit Keplerian model of the system is interpreted as a set of osculating initial conditions for an N-body dynamical system, the RMS to the fit jumps from $\sigma_{v}=2.3\,{\rm m\,s^{-1}}$ to $\sigma_{v}=4.4\,{\rm m\,s^{-1}}$ as a result of planet-planet interactions. These interactions are largely manifested as orbital precession of the inner planetary orbits. Optimizing a self-consistent N-body model of the system yields a system whose elements and goodness-of-fit are highly reminiscent of the Keplerian version, and the self-consistent N-body version of the system is the version listed in Table \ref{tab:fit}.

 \begin{figure}
 \plotone{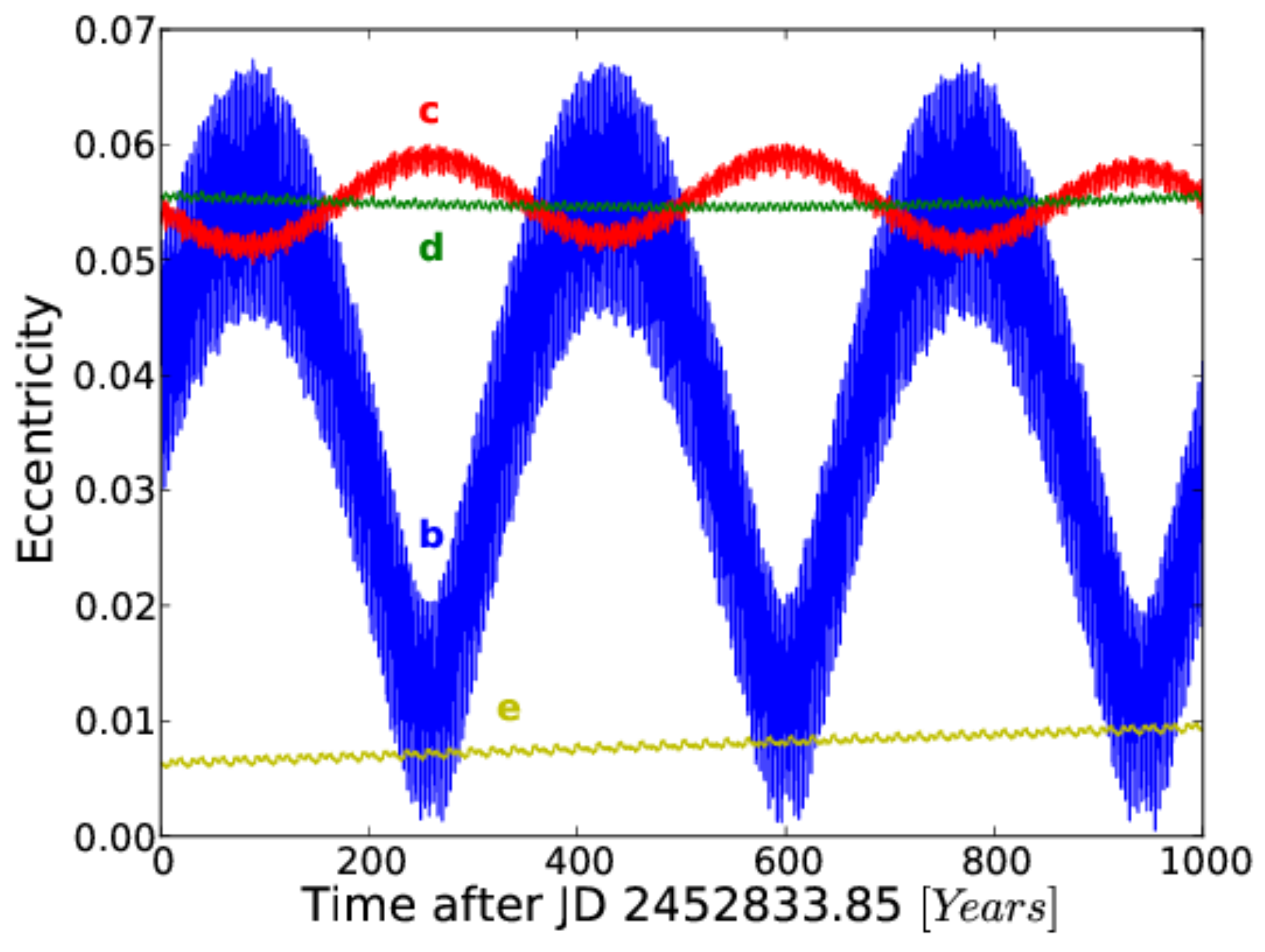}
 \caption{Eccentricity evolution for the 4-planet system listed in Table \ref{tab:fit} integrated for 1000 years. The eccentricity variations between planets b and c occur at a higher frequency than predicted by second-order secular perturbation theory, and likely stem from the contributions of resonant terms in the planetary disturbing function. When integrated, the system remains entirely stable through millions of years of further orbital evolution.}\label{fig:integration}
 \end{figure}

 As a zeroth-order evaluation of stability, we integrated the model of Table \ref{tab:fit} for 10 million years, observing stable behavior. Figure \ref{fig:integration} shows the change in eccentricity for each of the four planets over 1,000 years of this integration. Planets b and c engage in an exchange of eccentricities with a timescale of $P\sim250$ years. This relatively short timescale is at odds with the eigenfrequencies from the Laplace-Lagrange second-order secular theory, which corresponds to periods of several thousand years. The high-frequency exchange of angular momentum between b and c arises from the terms in the disturbing function associated with the near 2:1 commensurability, and is an analogue of the so-called ``Great Inequality'' between Jupiter and Saturn, which has a characteristic period of P $\sim$ 1000 years, and is the result of Jupiter and Saturn's proximity to a 2:5 orbital commensurability.

 The system configuration listed in Table \ref{tab:fit} is a single representative from the range of plausible models for the system. We have therefore drawn 100 models randomly from the converged segment of the Markov Chain that was used to generate our uncertainty estimates. These alternate models were each integrated for 1 million years. In all cases, dynamical stability was observed in the sense that semi-major axis and eccentricity variations were both small and periodic in nature. In all of the models tested, both 2:1 resonant arguments (for planets b and c) were circulating. In none of the cases were one or more of the 5:1 resonant arguments (for planets c and d) observed to be librating.

 In order to test the viability of system configurations that are substantially inclined to the line of sight, we tested co-planar versions of the system in which the planetary masses are increased by a factor $f=1/\sin(i)$, with $f$ ranging from unity to near zero. Instability (within 10 Myr) was observed to set in at $f_{\rm crit}=5.1$, indicating that the overall system inclination (assuming coplanarity) is $i>11.3^{\circ}$.

 \section{Photometric Observations}

 We used TSU's T3 0.4~m Automatic Photoelectric Telescope (APT) located
 at Fairborn Observatory to acquire nightly photometric observations of
 HD~141399 during its 2010--2013 observing seasons. We programmed the APT
 to observe HD~141399 ($V=7.21$, $B-V=0.77$, K0V) differentially with
 respect to comparison star HD~140612 ($V=7.01$, $B-V=0.37$, F0V)
 and check star HD~139798 ($V=5.76$, $B-V=0.35$, F2). Each observation was made in the following sequence, which we call a group observation:
 {\it K,S,C,V,C,V,C,V,C,S,K}, where $K$ is the check star, $C$ is the
 comparison star, $V$ is the target star, and $S$ is a sky reading. Three
 $V-C$ and two $K-C$ differential magnitudes are computed from each sequence
 and averaged into group means. Group mean differential magnitudes with
 internal standard deviations greater than 0.01 mag were rejected to eliminate observations taken under non-photometric conditions. The remaining group means were corrected for extinction and transformed to the Johnson $UBV$ system. The precision of a single group mean differential magnitude on a
 good night is usually in the range $\sim$0.003--0.006 mag
 \citep{Henry2000}, depending primarily on the brightness of the stars and the airmass of the observation. Additional information on the operation of the T3 APT can be found in \citet{Henry95a,Henry95b} and \citet{Eaton2003}.

 \begin{figure}
 \plotone{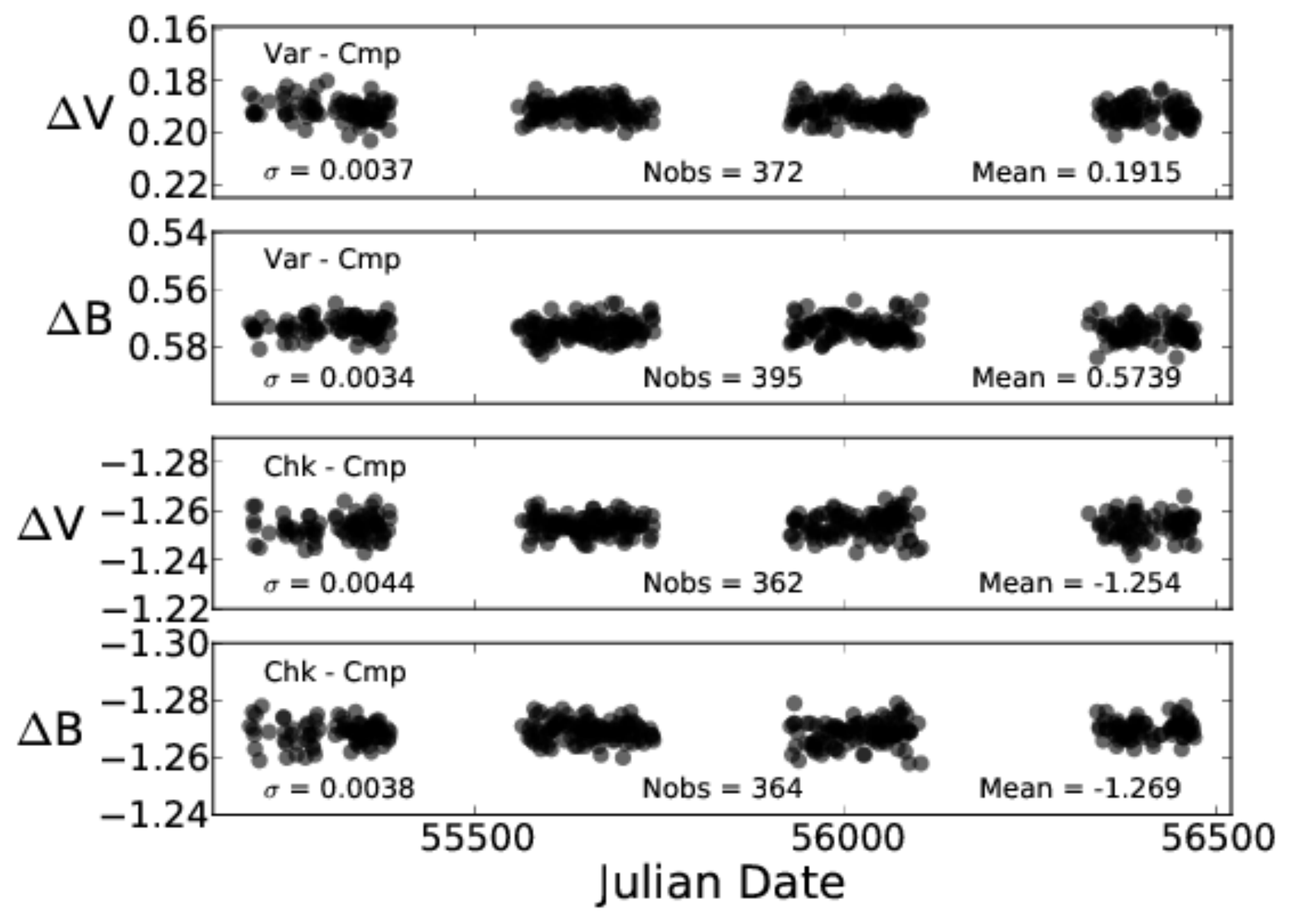}
 \caption{Comparitive photometric observations of HD~141399. Shown are the differential magnitudes in both the $V$ and $B$ bands for both the check star and comparison star. The groupings of points are the yearly groups of observations from 2010 to 2013.}
 \label{fig:phot}
 \end{figure}

 Photometric observations can be useful for eliminating potential false
 positives from radial velocity exoplanet surveys. For example,
 \citet{Queloz01} and \citet{Paulson04} have demonstrated how rotational
 modulation in the visibility of starspots on active stars can result in
 periodic radial velocity variations and, therefore, mimic the presence of
 a planetary companion. Our photometric results for HD~141399 are shown in Figure \ref{fig:phot}. The figure documents the total number of observations in each band, the mean of those observations, and their standard deviation. If we break down the observations by their respective observing year, the yearly means all agree closely, and their standard deviations range from 0.0004 to 0.0007 mag, indicating that the yearly mean magnitudes are stable to better than one millimagnitude from year to year. In addition, periodogram analysis of each data set found no significant periodicity between 1 and 100 days. We find no evidence for photometric variability to high precision in any of the three stars and conclude that stellar activity in HD~141399 has no significant effect on the radial velocity measurements \citep[see e.g.,][]{Boisse2012}.

 \section{Discussion}

 \begin{figure}
 \plotone{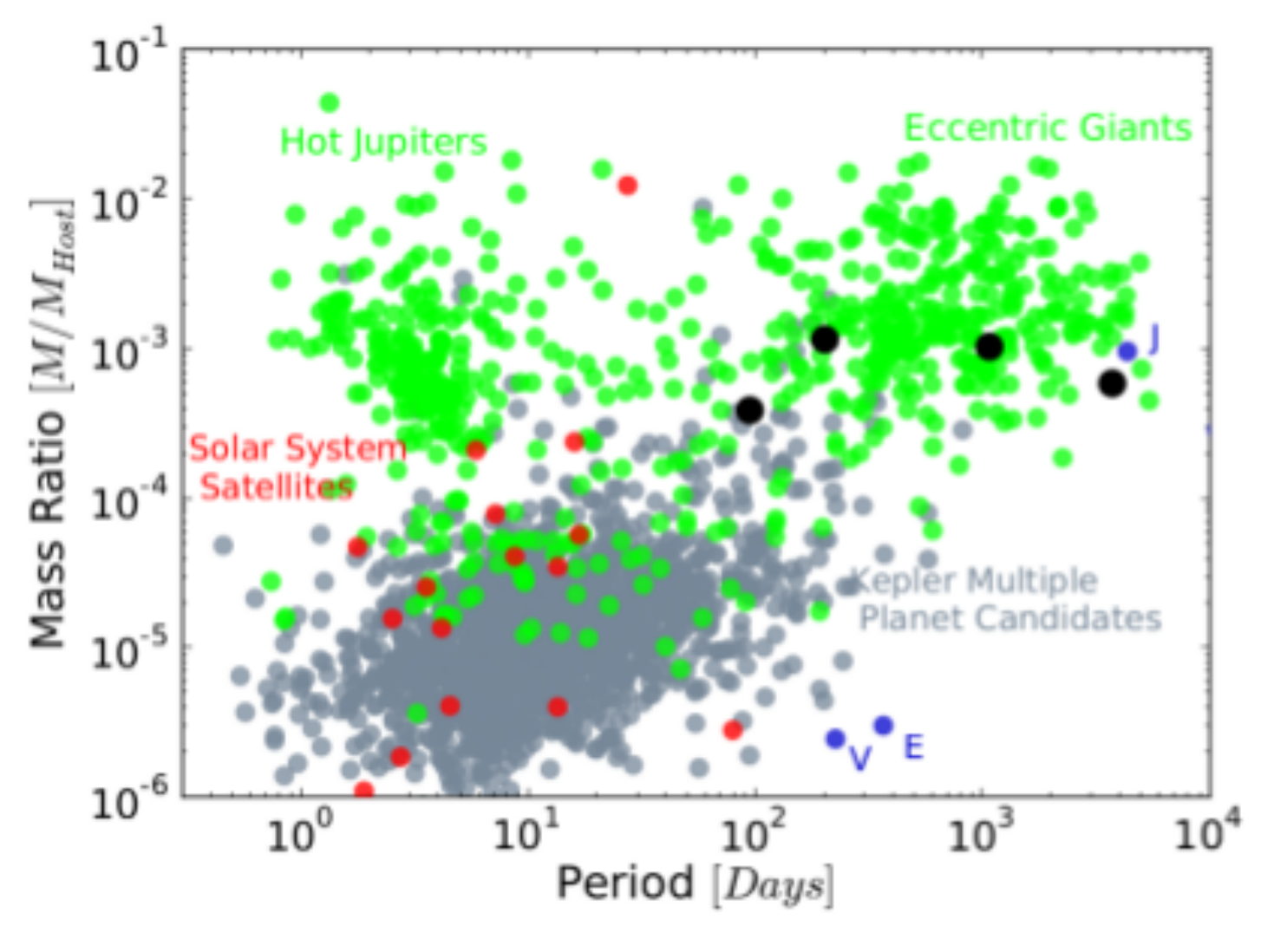}
 \caption{{\it Green circles:} 634 planets securely detected by the radial velocity method (either with or without photometric transits). {\it Red circles:} The regular satellites of the Jovian planets in the Solar System. {\it Gray circles:} 1501 Kepler candidates and objects of interest in which {\it multiple} transiting candidate planets are associated with a single primary. Radii for these candidate planets, as reported in \citep{Batalha13}, are converted to masses assuming $M/M_{\oplus}=(R/R_{\oplus})^{2.06}$ \citep{Lissauer11}, which is obtained by fitting the masses and radii of the solar system planets bounded in mass by Venus and Saturn. Data are from www.exoplanets.org and exoplanetarchive.ipac.caltech.edu, accessed 03/03/2013. \label{fig:period_mass_ratio}}
 \end{figure}

 \begin{figure}
 \plotone{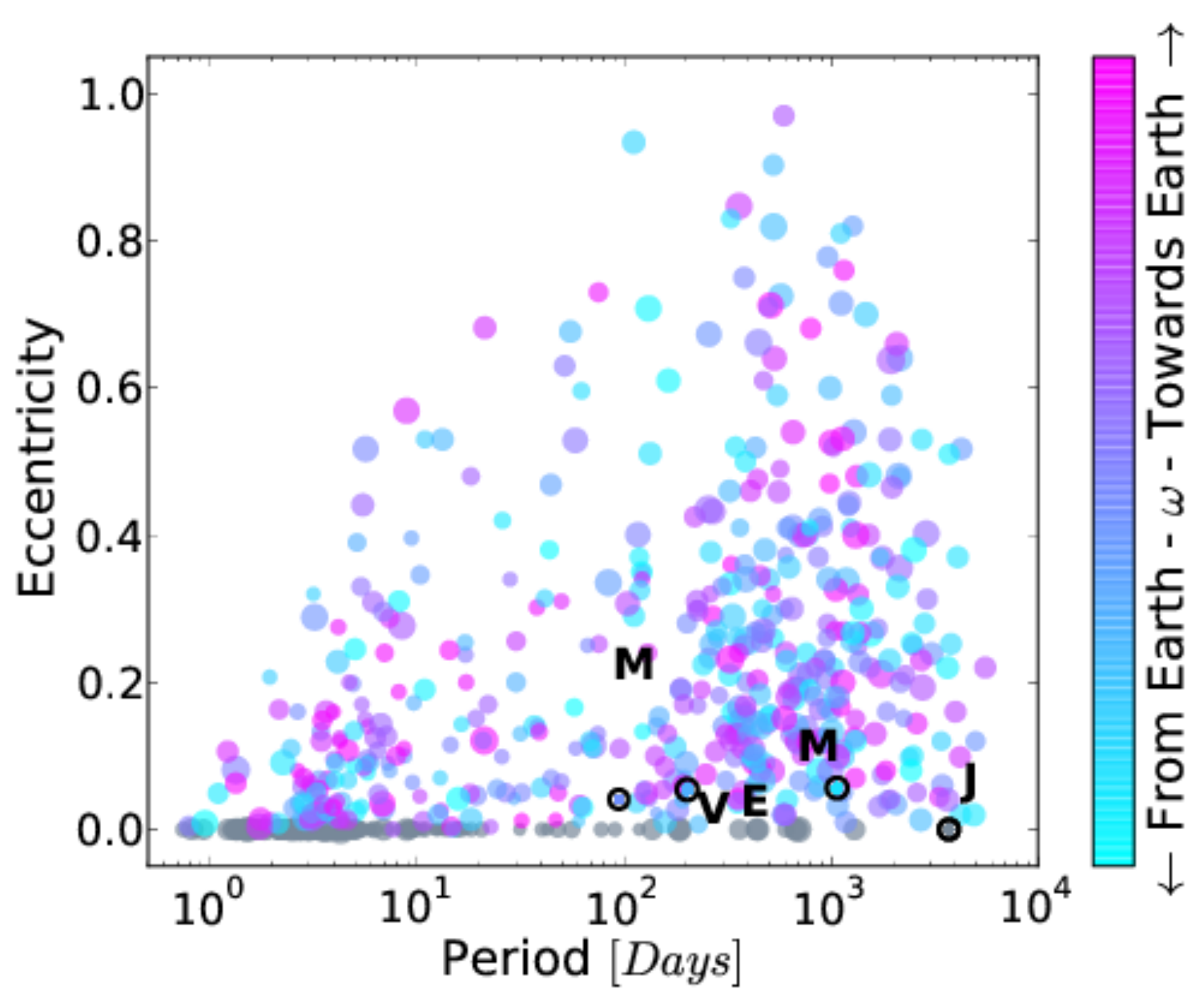}
 \caption{Eccentricity-Period diagram for extrasolar planets detected via Doppler Velocity measurements. HD 141399 b, c, d, and e are indicated as open circles, and the positions of the Solar System Planets are indicated as well. Symbol size is proportional to $(M\sin(i))^{1/3}$. The color scale represents angular separation between the longitude of periastron and the line of sight to the Earth. Magenta coloring indicates that the periastron line is close to the line of sight to Earth, and signals a higher probability of transit. Planets with circular orbits or unmeasured eccentricities are indicated in gray.}
 \label{fig:period_ecc}
 \end{figure}

 HD~141399's retinue of planets joins a relatively select group of systems that orbit bright parent stars and which also contain multiple Jovian-mass planets observed over multiple orbital periods. As with other members of this particular class, the orbital parameters can be determined with a high degree of precision. A number of such configurations, with prime examples being provided by Upsilon Andromedae \citep{Butler99}, 55 Cancri \citep{Marcy02}, and Gliese 876 \citep{Marcy98}, were discovered during the first decade of the high-precision Doppler velocity surveys, but the discovery rate of such systems has fallen off in recent years, as ease of detection has been supplanted by intrinsic scarcity. Indeed, we expect that relatively few additional systems with such unambiguously determined orbits and large planetary radial-velocity half-amplitudes remain to be discovered orbiting $V<8$ primaries.

 This system -- with three Jovian-mass planets lying at stellocentric distances normally associated with the terrestrial planets of our own solar system -- is fundamentally alien. As shown in Figure \ref{fig:period_mass_ratio} and Figure \ref{fig:period_ecc}, HD~141399's companions do, however, lie within the outskirts of the now well-demarcated distribution of giant planets that have been found over the years by the radial velocity surveys. It appears that $\sim10$\% of the F, G, and K dwarf stars in the solar neighborhood \citep{Cumming08} harbor such planets, although in many cases they are (i) more massive than Jupiter, (ii) of somewhat longer period than HD~141399 b and c, and (iii) single with large orbital eccentricity.

 It is of substantial interest to know whether planets such as the companions to HD~141399 formed {\it in situ}, or whether they accreted the bulk of their mass further out in the protoplanetary disk and subsequently suffered Type II migration and attendant orbital decay \citep{2011exop.book..347L}. In this regard, the proximity of planets b and c to the 2:1 mean motion resonance may provide an important clue. A history of quiescent inward migration would suggest that these two planets should have been captured into resonance. Their current configuration, however, can be placed outside of the resonance with a very high degree of confidence, due to the high precision to which the orbital eccentricities have been determined.

 The aggregate of candidate systems with multiple transiting planets observed by Kepler show no overall preference for configurations of planets lying in low order mean motion resonance. The Kepler systems do, however, show a mild preference for configurations in which the period ratios are a few percent larger than the nominal resonant value \citep{Lissauer11}. The HD 141399 system conforms to this particular pattern. Recently, \citet{2013AJ....145....1B} and \citet{2012ApJ...756L..11L} have shown that when two planets in the vicinity of a low-order resonance interact gravitationally in the presence of dissipation, the initial orbital separation increases as orbital energy is converted to heat. Initially near-resonant pairs are driven toward orbits that are both more circular and separated by an increased distance that scales with the total integrated dissipation experienced. \citet{2012ApJ...756L..11L} suggest that the observed overdensity of near-resonant pairs can arise if tidal dissipation is unexpectedly efficient, with $Q\sim10$. This explanation seems unlikely for HD 141399 b and c, which are likely gas giants, and which likely have tidal quality factors that are orders of magnitude away from the required value. \citet{2013AJ....145....1B} argue that the dissipative mechanism is provided by interaction with the surrounding protoplanetary disk. This mechanism would appear to be more viable in this case, although the hydrodynamical details are somewhat vague and remain to be worked out.

 In conclusion, HD~141399 harbors a fairly unusual system in which three (and likely a fourth) Jovian-mass planets lie on low-eccentricity orbits with periods that conform to one's naive expectation for terrestrial planets. Confirmation of this system was significantly aided by velocity measurements from the APF telescope. The quality of the measurements that the APF is obtaining show it is functioning as intended, and that it is producing Doppler measurements that conform to the current state-of-the-art.

 \acknowledgments
 GL acknowledges support from the NASA Astrobiology Institute
 through a cooperative agreement between NASA Ames Research Center and the University
 of California at Santa Cruz. SSV gratefully acknowledges support from
 NSF grants AST-0307493 and AST-0908870. RPB gratefully acknowledges support from NASA OSS
 Grant NNX07AR40G, the NASA Keck PI program,
 and from the Carnegie Institution of Washington. S.M. acknowledges support from the W. J.
 McDonald Postdoctoral Fellowship.
 GWH acknowledges support from NASA, NSF, Tennessee State University, and
 the State of Tennessee through its Centers of Excellence program.
 The work herein is based on
 observations obtained at the W. M. Keck Observatory, which is operated jointly by the University of California and
 the California Institute of Technology, and we thank the UC-Keck, UH, and NASA-Keck Time Assignment Committees for their
 support. This research has made use of the Keck Observatory Archive (KOA), which is operated
 by the W. M. Keck Observatory and the NASA Exoplanet Science Institute (NExScI), under contract with
 the National Aeronautics and Space Administration. We thank those who collected the data in the KOA
 including: D. Fischer, G. Marcy, J. Winn, W. Boruki, G. Bakos, A. Howard, and T. Bida. We also wish to extend our
 special thanks to those of Hawaiian ancestry on whose sacred mountain
 of Mauna Kea we are privileged to be guests. Without their generous hospitality, the Keck observations
 presented herein would not have been possible. This research has made use of the SIMBAD database,
 operated at CDS, Strasbourg, France. This paper was produced using $^BA^M$.

 \vspace{\baselineskip}
 {\it Facilities:} \facility{Keck (HIRES)} \facility{Automated Planet Finder (Levy Spectrometer)}.

 \clearpage
 \bibliographystyle{apj}
 \bibliography{biblio}

 \end{document}